\documentclass[aps,prd,nofootinbib,english, 11pt]{revtex4-2}
\usepackage[utf8]{inputenc}

\usepackage{graphicx}
\usepackage[dvipsnames]{xcolor}
\usepackage{rotating}
\usepackage{amssymb}
\usepackage{amsfonts}
\usepackage{amsmath}
\usepackage{xspace}
\usepackage{mathtools}
\usepackage{verbatim}
\usepackage{comment}
\usepackage{enumitem}
\usepackage{physics}
\usepackage{graphicx}
\usepackage[colorlinks]{hyperref}
\usepackage{MnSymbol}
\usepackage{graphicx}
\usepackage{accents}
\usepackage{soul}

\begin{document}

\title{Cosmological Landsberg Finsler spacetimes}

\author{Annamária Friedl-Szász}
\email{szasz.annamaria@unitbv.ro}
\affiliation{Faculty of Mathematics and Computer Science, Transilvania University, Iuliu Maniu Str. 50, 500091 Brasov, Romania}

\author{Christian Pfeifer}
\email{christian.pfeifer@zarm.uni-bremen.de}
\affiliation{ZARM, University of Bremen, 28359 Bremen, Germany and Faculty of Mathematics and Computer Science, Transilvania University, Iuliu Maniu Str. 50, 500091 Brasov, Romania}

\author{Elena Popovici-Popescu}
\email{popovici.elena@unitbv.ro}
\affiliation{Faculty of Mathematics and Computer Science, Transilvania University, Iuliu Maniu Str. 50, 500091 Brasov, Romania}

\author{Nicoleta Voicu}
\email{nico.voicu@unitbv.ro}
\affiliation{Faculty of Mathematics and Computer Science, Transilvania University, Iuliu Maniu Str. 50, 500091 Brasov, Romania}

\author{Sjors Heefer}
\email{s.j.heefer@tue.nl}
\affiliation{Department of Mathematics and Computer Science, Eindhoven University of Technology, Eindhoven 5600MB, The Netherlands}

\begin{abstract}
We classify all possible cosmological homogeneous and isotropic Landsberg-type Finsler structures, in 4-dimensions. Among them, we identify viable non-stationary Finsler spacetimes, i.e. those geometries leading to a physical causal structure and a dynamical universe. Noting that any non-stationary and non-Riemannian Landsberg metric must be actually also non-Berwald (i.e., it should be a so-called \textit{unicorn}), we  construct the \emph{unique} Finsler, non-Berwaldian Landsberg generalization of Friedmann-Lemaitre-Robertson-Walker geometry.
\end{abstract}

\maketitle


\section{Introduction}
The gravitational fields of physical systems that are described in terms of kinetic gases, can be understood from a new perspective, namely in terms of Finsler gravity \cite{Hohmann:2020yia,Hohmann:2019sni}.

In the standard (Einstein-Vlasov \cite{Andreasson:2011ng,Sarbach:2013fya}) description of the gravitational field of a kinetic gas in general relativity (GR), a lot of the information of the gas is lost.

More precisely, the kinematics and dynamics of multi-particle systems described as a kinetic gas are encoded into the 1-particle distribution function (1PDF) $\varphi$, which lives, depending on the formulation, on the tangent or on the cotangent bundle (i.e., the space of all positions and velocities, respectively, the space of all positions and momenta of particles) of spacetime. The 1PDF contains all the information about the kinetic gas. In general relativity (GR), the gravitational field of kinetic gases is derived from the Einstein equations, coupled to an energy-momentum tensor that is obtained from the 1PDF by averaging over the normalized 4-velocities in the following way:
\begin{align}
    T^{ab}_{KG} = \int_{\mathcal{S}_x} d\Sigma_x \frac{\dot x^a \dot x^b}{g(\dot x, \dot x)} \varphi(x,\dot x)\,,
\end{align}
where $\mathcal{S}_x$ is the set of normalized $4$-velocities $\dot x$ and $d\Sigma_x$ is a suitable volume form \cite{Ehlers2011,Sarbach:2013uba,Sarbach:2013fya,Andreasson:2011ng}. Using this kind of coupling between the kinetic gas and the geometry of spacetime, only specific aspects of the kinetic gas are taken into account in the derivation of its gravitational field, namely the second moments of the 1PDF with respect to the 4-velocities of the gas particles.

Two immediate questions arise immediately:
\begin{itemize}
    \item How do all other moments of the 1PDF
    \begin{align}
    \mathfrak{T}^{a_1...a_n}(x) = \int_{\mathcal{S}_x} d\Sigma_x \frac{\dot x^{a_1}....\dot x^{a_n}}{\sqrt{g(\dot x, \dot x)}^n} \varphi(x,\dot x)\,,
    \end{align}
    which certainly contain non-trivial information about the kinetic gas, contribute to its gravitational field? Why are they neglected in the coupling to the Einstein equations?
    \item How much information is lost in the averaging procedure?
\end{itemize}

Finsler gravity offers the opportunity to answer these questions. Finsler geometry is a straightforward generalization of (pseudo-)Riemannian geometry \cite{Finsler, Bucataru, BCS}, based on a general geometric length measure for curves. It describes the geometry of spacetime in terms of geometric objects (a canonical nonlinear connection and its curvature) that naturally live on the tangent bundle of spacetime. There exist various applications of Finsler geometry in physics \cite{Pfeifer:2019wus}. In particular, it leads to a natural extension of general relativity, generally known as Finsler gravity \cite{Rutz,Triantafyllopoulos:2020ogl,Villasenor:2022-1,Garcia-Parrado:2022ith}. It turns out that the action-based canonical formulation of Finsler gravity introduced in \cite{Pfeifer:2011tk, Hohmann_2019} can very naturally be coupled to the full 1PDF of the kinetic gas \cite{Hohmann:2020yia,Hohmann:2019sni,Pfeifer:2019wus}. This leads to a canonical variational Finsler gravity equation, of the the form:
\begin{align}\label{eq:Fgrav}
    \mathcal{G}(x,\dot x) := \mathcal{R}(x,\dot x) + \mathcal{P}(x,\dot x) = \kappa \varphi(x,\dot x)\,,
\end{align}
where $\mathcal{R}=\mathcal{R}(x,\dot x)$ and $\mathcal{P}=\mathcal{P}(x,\dot x)$ are scalar functions built from the Finsler curvature tensor, respectively, from a Finslerian quantity called the Landsberg tensor, and $\kappa$ is a constant (to be determined).

The goal of Finsler cosmology---the study of Finsler gravity in homogeneous and isotropic symmetry---is to describe the evolution of the universe in accordance with observation, without the need for $95\%$ of dark (matter and energy) constituents of the universe. The conjecture is that the gravitational field of the neglected parts of the 1PDF of the cosmological kinetic gas account for (at least parts of) the effects that are typically associated with dark matter and dark energy, when their contribution to the geometry of spacetime is properly taken into account \cite{Hohmann:2020yia}. Independently of this conjecture relating dark matter and dark energy to properties of kinetic gases and canonical action-based Finsler gravity, there exist other approaches to relate dark matter and dark energy to Finsler geometry \cite{Kouretsis:2008ha,Mavromatos:2010nk,Papagiannopoulos:2017whb,Li:2015sja,Hohmann:2016pyt,CANTATA:2021asi,Kapsabelis:2023khh,Stavrinos:2024noe}.

In general, the above Finsler gravity equation is difficult to solve and so far, only several solutions have been found \cite{Fuster:2015tua,Fuster:2018djw,Heefer_2021,Heefer_2023_Finsler_grav_waves,Heefer:2023a}, see \cite{Heefer:2024kfi} for an overview. A particular difficulty in solving the Finsler gravity equation is introduced by the $\mathcal{P}$-term, involving the Landsberg tensor. Intuitively, the Landsberg tensor measures the rate of change of the `non-Riemannianity' of the geometry as one moves along geodesics of spacetime. It is natural to first look for solutions to the Finsler gravity equation whose Landsberg tensor vanishes (called \emph{Landsberg spacetimes}) because in that case, the equation simplifies significantly and the departure from pseudo-Riemannian geometry is in some sense mild. Landsberg manifolds admit a particular subclass, called \emph{Berwald manifolds},  \cite{Berwald1926}, for which the canonical (in general nonlinear) Finsler connection is a linear (affine) connection on spacetime. These are the simplest Finsler manifolds, closest to pseudo-Riemannian geometry, and the most investigated ones in the literature \cite{Szabo,Crampin,Fuster:2018djw,Hohmann:2020mgs,Fuster:2020upk,Pfeifer:2019tyy,Cheraghchi:2022zgv,Voicu:2023fey,Fuster:2015tua,Heefer_2021,Heefer_2023_Finsler_grav_waves, Heefer2023mKropNull, heefer2024berwaldmkropinaspacesarbitrary}.  The various classes of Finsler spacetimes can be summarized concisely by their departure from pseudo-Riemannian geometry as follows:
\begin{align}\label{eq:classes}
	\textrm{pseudo-Riemannian} \subset \textrm{Berwald} \subset \textrm{Landsberg} \subset \textrm{general Finsler}\,.
\end{align}
Importantly however, in the presence of cosmological symmetry it was proven that Berwald structures are too simple to describe the evolution of the universe beyond general relativity, as any solution to the field equation of this type must be either a pseudo-Riemannian Friedmann-Lemaitre-Robertson-Walker (FLRW) geometry, or a stationary one \cite{Hohmann:2020mgs}. Thus, the next more interesting class to be investigated is the class of Landsberg spacetimes which are non-Berwaldian. Since these are notoriously difficult to find (actually, a full classification thereof is an important open problem in mathematics, \cite{Elgendi2021a,Elgendi2021b,elgendi_2020,Bao_unicorns}), non-Berwaldian Landsberg space(time)s have been called \emph{unicorns} \cite{Bao_unicorns}.

In this article, we classify all $4$-dimensional homogeneous and isotropic Finsler structures of Landsberg type (in particular, all those that are Landsberg, but non-Berwald).

Furthermore, we prove that among these, there is a \emph{unique} class that admits a well-defined causal structure (featuring both a well-defined, convex future-pointing timelike cone and a past one, at each point) and hence represents a direct Finsler, non-Berwaldian Landsberg generalization of 
FLRW geometry. The obtained Landsberg-type generalization of the FLRW metric can then be used as an ansatz in the Finsler gravity equation, to derive the Finsler gravity analogue of the Friedmann equations (however, this is a task for the future). Moreover, our findings solve the mathematical question of the full classification of non-Berwaldian Landsberg Finsler geometries in $4$ dimensions and under the assumption of homogeneous and isotropic symmetry.

In order to derive our main result---the identification of the non-Berwald Landsberg generalization of FLRW geometry (in equation \eqref{eq:FinslerUnicornFLRW})---we begin this article by recalling the necessary mathematical concepts of Finsler geometry, the different classes (Berwald, Landsberg and weakly Landsberg) of Finsler manifolds, and the form of homogeneous and isotropic Finsler spacetime functions in Section \ref{sec:prelims}. Afterwards, in Section \ref{sec:Unicorn}, we derive and solve the Landsberg condition and thus find all possible $4$-dimensional homogeneous and isotropic non-Berwald Landsberg Finslerian manifolds. Among them, we identify the only one which qualifies as a Finsler spacetime in Section \ref{sec:UnicornSpacetime}, after which we discuss our results and conclude in Section \ref{sec:disc}.

\section{Preliminaries - Homogeneous and isotropic Finsler spacetime geometry}\label{sec:prelims}
This section introduces the mathematical preliminaries to Finsler spacetime geometry and its application to cosmological homogeneous and isotropic symmetry. We start by introducing Finsler spacetimes and Finsler geometry in Subsection \ref{ssec:FGeom}. Then, we discuss different classes of Finsler spacetimes and identify the relevant non-Berwald Landsberg type Finsler manifolds in Subsection \ref{ssec:finsClass} before we recall homogeneous and isotropic symmetry in Finsler geometry in Subsection \ref{ssec:homiso}.

Throughout this article, we consider $M$ to  be a 4-dimensional connected, orientable smooth manifold, $(TM,\pi,M)$, its tangent bundle and $\overset{\circ}{TM}=TM\backslash\{0\}$ the so-called slit tangent bundle, i.e.\ the tangent bundle without its zero section. We will denote by $(x^a)_{a=\overline{0,3}}$, the coordinates of a point $x\in U\subset M$ in a local chart $\left(U,\varphi\right)$ and by $(x^a,\dot{x}^a)$, the naturally induced local coordinates of points $(x,\dot{x})\in\pi^{-1}(U)$ in the tangent bundle. Also, whenever there is no risk of confusion, we will omit for simplicity the indices of the coordinates. Moreover we introduce the following shorthand notations for derivatives:
\begin{align}
	\partial_a = \frac{\partial}{\partial x^a}\,,\quad \dot \partial_a = \frac{\partial}{\partial \dot x^a}\,.
\end{align}

Indices $a,b,c,...$ will run over all spacetime dimensions $0,1,2,3$ while indices $\alpha, \beta, ...$ will run over the spatial dimensions $1,2,3$.

\subsection{Finsler spacetime geometry }\label{ssec:FGeom}
Finsler spacetimes are a straightforward generalization of pseudo-Riemannian spacetimes, where the geometry, and hence the description of gravity, is derived from a pseudo-norm instead of a spacetime metric.

Following \cite{math-foundations,Hohmann:2019sni}, let $\mathcal{A}\subset TM \backslash \{0\}$ be a conic subbundle of $TM$, meaning that $\pi (\mathcal{A}) = M$ and
\begin{align}
	(x,\dot x)\in \mathcal{A} \Leftrightarrow (x,\lambda \dot x)\subset \mathcal{A}.
\end{align}
A \emph{Finsler spacetime} $(M,L)$ is a manifold $M$ equipped with a  smooth, positively $2$-homogeneous Finsler Lagrangian~$L$
\begin{align}
	L: \mathcal{A}\rightarrow \mathbb{R}\,, \quad L(x,\lambda \dot x)= \lambda^2 L(x,\dot x), \quad \forall \lambda >0,
\end{align}
such that:
\begin{itemize}
    \item the \emph{Finsler metric tensor} $g_{ab}(x,\dot x)$, given by:
    \begin{align}\label{eq:fmet}
        g_{ab}(x,\dot x):= \frac{1}{2}\dot{\partial}_a \dot{\partial}_bL(x,\dot x)\,,
    \end{align}
    is nondegenerate on $\mathcal{A}$;
    \item there exists a smooth conic subbundle $\mathcal{T}\subset \mathcal{A}$ with connected fibers $\mathcal{T}_x = \mathcal{T}\cap T_xM$,  $\forall x\in M$, such that on each $\mathcal{T}_x$ we have:  $L>0$, $g_{ab}$ has Lorentzian signature $(+,-,-,-)$ and $L$ can be continuously prolonged as 0 to the boundary $\partial\mathcal{T}_x$ of $\mathcal{T}_x$.
\end{itemize}
This definition of Finsler spacetimes ensures the existence of the following physically necessary structures:
\begin{itemize}
    \item A large enough set  of directions $\mathcal{A}$ along which the needed geometric objects are well defined.
    \item A set of non-trivial null directions $\mathcal{N}:=\{(x,\dot x)\in TM | L(x,\dot x)=0\}$, interpreted as directions along which light propagates.
    \item A set $\mathcal{T}$ of convex cones of past or future pointing timelike directions.
    \item A set $\mathcal{S}:=\{(x,\dot x)\in \mathcal{T} | L(x,\dot x)=1\}$, of unit normalized directions in $\mathcal{T}$.
\end{itemize}
These sets generalize the corresponding structures that are defined through the spacetime metric on pseudo-Riemannian spacetimes.

Moreover, on a Finsler spacetime, we have the following parametrization-invariant\footnote{By  parameterization invariance we mean here invariance under \emph{orientation-preserving} changes of the parameter.} length measure for curves $\gamma$:
\begin{align}\label{eq:flength}
	S[\gamma] = \int d\tau \sqrt{|L(\gamma, \dot \gamma)|}\,.
\end{align}
For timelike curves, $\dot \gamma \in \mathcal{T}_{\gamma}$, this length is physically interpreted as the proper-time measure, or geometric clock. In this sense, Finsler spacetimes are spacetimes whose geometry is based on \emph{general} geometric clocks. The $1$-homogeneous function $F(x,\dot x) = \sqrt{|L(\gamma, \dot \gamma)|}$, is called the \emph{Finsler function} and is the fundamental building block in the classical literature on positive definite Finsler geometry \cite{BCS,Bucataru}.

When passing from positive definite to indefinite Finsler geometry, there exist various definitions of Finsler spacetimes by different authors, which all differ slightly depending on the application the authors are interested in \cite{Beem,Pfeifer:2011tk,Bernal2020,Lammerzahl:2012kw,Hasse:2019zqi,Caponio-Masiello,Caponio-Stancarone}. Their main difference lies in the assumption on where the Finsler Lagrangian is smooth and where the Finsler metric is nondegenerate. We will not enter the discussion on Finsler spacetimes here, and refer the interested reader to the references and the overview article \cite{math-foundations}. For the purposes of the analysis of homogeneous and isotropic Finsler gravity, we will work with the definition presented in this section.

The geometry of Finsler spacetimes is derived from the Finsler Lagrangian, in a very similar way as pseudo-Riemannian spacetime geometry is derived from the spacetime metric, see for example \cite{BCS,Bucataru}. Relevant geometric notions in Finsler geometry, besides the Finsler metric \eqref{eq:fmet}, that we will need throughout this article are:
\begin{itemize}
    \item The Cartan tensor and its trace:
	\begin{align}\label{eq:Cartan}
		C_{abc} = \frac{1}{4}\dot{\partial}_a \dot{\partial}_b \dot{\partial}_c L = \frac{1}{2} \dot{\partial}_c g_{ab}, \quad  C_a = g^{bc}C_{abc}\,.
	\end{align}
        The Cartan tensor measures, at each point $x \in M$, how much the Finslerian metric tensor $g_{a,b}(x,\dot{x})$ differs from a pseudo-Riemannian one.
    \item The geodesic spray coefficients $G^a$, the canonical nonlinear connection coefficients $N^a{}_b$, and the horizontal derivatives $\delta_a$:
	\begin{align}\label{eq:geodspray}
		G^a = \frac{1}{4} g^{ab}(\dot x^c \partial_c \dot \partial_b L - \partial_b L)\,,\quad N^a{}_b = \dot{\partial}_b G^a\,,\quad \delta_a = \partial_a - N^b{}_a \dot\partial_b\,.
	\end{align}
        The geodesic spray defines the geodesic equation from the Euler Lagrange equations of the Finsler length \eqref{eq:flength} in arclength parameterisation.
        \item The curvature of the canonical nonlinear connection and the canonical curvature scalar (or \textit{Finsler-Ricci} scalar:
	\begin{align}\label{eq:curv}
		R^c{}_{ab}\dot \partial_c = [\delta_a,\delta_b]\,,\quad R = R^a{}_{ab}\dot x^b \,.
	\end{align}
        \item The Chern-Rund covariant derivatives, defined on the \emph{adapted basis}  $\{\delta_a,\dot{\partial}_a\}$ of $T_{(x,\dot x)}TM$ as:
	\begin{align}
		\nabla_{\delta_{a}}\delta_{b}&=\Gamma^{c}{}_{ab}\delta_{c}=\frac{1}{2}g^{cd}\left(\delta_{a}g_{bd}+\delta_{b}g_{ad}-\delta_{d}g_{ab}\right)\delta_{c}\,, &  & \nabla_{\delta_{a}}\dot{\partial}_{b}=\Gamma^{c}{}_{ab}\dot{\partial}_{c}\,,\\
		\nabla_{\dot{\partial}_{a}}\delta_{b}&=0, &  & \nabla_{\dot{\partial}_{a}}\dot{\partial}_{b}=0\,.
	\end{align}
        \item The dynamical covariant derivative - measuring the rate of variation of tensor fields along geodesics of $(M,L)$,  is obtained from the Chern-Rund one by contraction with $\dot x^a$ and expressed by the symbol $\nabla:=\dot x^a \nabla_{\delta_{a}}$.
	\item The Landsberg tensor and its trace
	\begin{align}\label{eq:lands}
		P_{abc} = \nabla C_{abc}\,,\quad P_a = g^{bc}\nabla C_{abc}\,.
	\end{align}
       They keep track of how much the "non-Riemannianity" of our Finsler space varies as one moves along its geodesics.
\end{itemize}
Having clarified these notions, we can classify Finsler geometries according to how much more general they are compared to pseudo-Riemannian geometry.

\subsection{Classes of Finsler geometries}\label{ssec:finsClass}
In general, Finsler geometries are widely more general than pseudo-Riemannian ones. The latter are defined in terms of the ten component functions of a metric tensor on a smooth manifold, whereas the former are given by a $2$-homogeneous function on the tangent bundle --- which, from the viewpoint of the base manifold, has infinitely many degrees of freedom. Hence, it is fair to say that Finsler geometry gives us an enormous amount of additional freedom, compared to pseudo-Riemannian geometry.

Among all possible Finsler geometries, there are subclasses which are closer to pseudo-Riemannian geometry than others. We briefly recall them here in a ladder-structured way, from particular to general. The classification below holds in arbitrary signature, even if we only focus here on Finsler spacetimes:
\begin{itemize}
    \item Pseudo-Riemannian manifolds: A Finsler spacetime $(M,L)$ is pseudo-Riemannian  if and only if its Cartan tensor \eqref{eq:Cartan} vanishes, which means that the Finsler metric tensor \eqref{eq:fmet} is independent of $\dot x$, i.e. it is a pseudo-Riemannian metric.
    \item Berwald pseudo-Finsler spaces $(M,L)$ (Berwald spaces): These are Finsler spaces whose Cartan tensor \eqref{eq:Cartan}  is non-vanishing, but whose nonlinear connection \eqref{eq:geodspray} is linear in $\dot x$:
    \begin{align}
	N^a{}_b(x,\dot x) = \Gamma^a{}_{bc}(x)\dot x^c;
    \end{align}
    equivalently, the canonical nonlinear connection arises from an affine connection with coefficients $\Gamma^a{}_{bc}(x)$ on the base manifold $M$ \cite{Berwald1926}. Berwald-Finsler functions $L$ can be found by fixing a (symmetric) affine connection on spacetime and solving the equation
    \begin{align}
	\delta_a L=\partial_a L - \Gamma^c{}_{ab}(x)\dot x^b\dot\partial_c L = 0\,.
    \end{align}
    Berwald geometries are sometimes addressed as the non-trivial Finsler geometries closest to pseudo-Riemannian geometry, and have been studied in depth from different angles, in both mathematics and physics \cite{Szabo,Crampin,Fuster:2018djw,Hohmann:2020mgs,Fuster:2020upk,Pfeifer:2019tyy,Cheraghchi:2022zgv,Voicu:2023fey,Fuster:2015tua,Heefer_2021,Heefer_2023_Finsler_grav_waves, Heefer2023mKropNull, heefer2024berwaldmkropinaspacesarbitrary}.
    \item Landsberg and weakly Landsberg geometries: A Finsler spacetime $(M,L)$  is called Landsberg, respectively, weakly Landsberg, if its Lansberg tensor \eqref{eq:lands}, respectively, its trace, vanishes
    \begin{align}
	P_{abc} = 0\,,\qquad P_a=0\,.
    \end{align}
    Landsberg geometries are those for which the  Cartan tensor is preserved by parallel transport along  geodesics, i.e. the deviation from pseudo-Riemannian geometry `looks the same' in each tangent space. In weakly Landsberg geometries only the canonical volume form is preserved. From equation \eqref{eq:lands} and $g^{ab} C_{abc}=\dot{\partial}_c \ln(\sqrt{|\det(g)|})$ it follows that $\nabla \dot \partial_c(\tfrac{|\det(g)|}{L^2})=0$.

    Weakly Landsberg and Landsberg geometries play an important role in Finsler gravity since, for this class the gravitational field equations simplify tremendously \cite{Hohmann_2019,Heefer:2023a}, and it has been shown that the Palatini and the metric formulation of action based Finsler gravity coincide \cite{Villasenor:2022-1}.
	
    It was believed for a long time that all Landsberg spaces were automatically Berwald. Landsberg spaces that are not Berwald turned out to be extremely difficult to find, to the point it was believed they might not exist at all; this is why the term \textit{unicorn geometries} was coined for them~\cite{Bao_unicorns}. In 2006, however, some unicorns were found by Asanov \cite{asanov_unicorns} and his results were generalized by Shen \cite{shen_unicorns} a few years later. Just very recently, a systematic way of constructing unicorns by conformal rescaling of Berwald metrics was introduced in~\cite{elgendi_2020,Elgendi2021a,Elgendi2021b}.
\end{itemize}

The just presented classes of Finsler geometries can nicely be summarized by the following inclusions:
\begin{align}
	\textrm{pseudo-Riemannian} \subset \textrm{Berwald} \subset \textrm{Landsberg} \subset \textrm{weakly Landsberg} \subset \textrm{general Finsler}\,.
\end{align}

Since we are aiming to derive all possible homogeneous and isotropic Landsberg geometries in 4-dimensions, we now briefly recall the most important known results about homogeneous and isotropic Finsler structures.

\subsection{Homogeneous and isotropic symmetry}\label{ssec:homiso}
The Copernican and cosmological principle leads us to the conclusion that, on largest scales, the universe is spatially homogeneous and isotropic \cite{Weinberg}. In four dimensions (which is what we assume for realistic spacetime geometries), this implies the existence of a global time function $t:M \rightarrow \mathbb{R}$ and of six Killing vector fields, not only in the pseudo-Riemannian case but also in the general Finsler setting \cite{Hohmann:2020mgs}.

More precisely, on each spatially homogeneous and isotropic 4-dimensional Finsler manifold there exist local coordinates $(t,r,\theta,\phi)$ such that the vector fields
\begin{subequations}\label{eq:csym}
	\begin{eqnarray}
		X_1&=&\chi\sin\theta \cos\phi
		\partial_r+\frac{\chi}{r}\cos\theta\cos\phi\partial_\theta-\frac{\chi}{r}\frac{\sin\phi}{\sin\theta}\partial_\phi\,,\\
		X_2&=&\chi\sin\theta \sin\phi
		\partial_r+\frac{\chi}{r}\cos\theta\sin\phi\partial_\theta+\frac{\chi}{r}\frac{\cos\phi}{\sin\theta}\partial_\phi\,,\\
		X_3&=&\chi\cos\theta \partial_r-\frac{\chi}{r}\sin\theta\partial_\theta\,,\\
		X_4&=&\sin\phi\partial_\theta+\cot\theta\cos\phi\partial_\phi\,,\quad
		X_5=-\cos\phi\partial_\theta+\cot\theta\sin\phi\partial_\phi\,,\quad
		X_6=\partial_\phi\,,
	\end{eqnarray}
\end{subequations}
generate diffeomorphisms on $M$ which leave the Finsler Lagrangian $L$ invariant. Here $\chi = \sqrt{1 - k r^2}$ and $k$ is the scalar curvature of the spatial slices $t=const$. Technically, this means that $L$ is constant along the flows of complete lifts of the above vector fields $X^C_I, I=1, ..6$:
\begin{align}
	X^C_I(L) = 0\,.
\end{align}
The complete lift of a vector field $\xi = \xi^a\partial_a$ on $M$ to $TM$ is given by $\xi^C = \xi^a\partial_a + \dot x^c \partial_c \xi^a \dot \partial_a$. It can now straightforwardly be shown that the Finsler Killing equations for $X_1$ to $X_6$ are satisfied if and only if \cite{Pfeifer:2011xi}
\begin{align}\label{eq:homisoL}
	L(t,r,\theta,\phi,\dot t, \dot r, \dot \theta, \dot \phi) = L(t,\dot t, w)\,,\quad w^2 = \frac{\dot r^2}{1-k r^2}+r^2\big(\dot \theta^2+\sin^2(\theta) \dot\phi^2\big)\,.
\end{align}
The $2$-homogeneity of the Finsler Lagrangian w.r.t. the $\dot x$ variables allows us to write down the most general homogeneous and isotropic Finsler Lagrangian in the following, form based on a function $h(t,s)$ of just two variables,
\begin{align}\label{eq:homisoL2}
    L(t,\dot t, w) = \dot t^2 L(t,1,s) =: \dot t^2 h(t,s)^2,\quad s=s(\dot t, x^\alpha, \dot x^\alpha)=\frac{w}{\dot t}\,,
\end{align}
which is the most convenient form to derive the Finsler geometric objects (this is done in Appendix \ref{app:HomAndIsoGeom})---in particular, the Landsberg tensor.
Thus, cosmological Finsler geometries have a very specific dependence on the coordinates. Actually, each spatial slice $\Sigma_t: t=const.$ is a 3-dimensional Riemannian manifold of constant curvature $k$ with metric given by $w^2$ - just as in the FLRW case. The difference is that in the Finslerian case, the intertwining between the time coordinate and the spatial part $w$ of the Finsler Lagrangian is much more general. 

The above finding on the geometry of the spatial slices also sheds light on the topological structure of $M$. In particular, there exists a diffeomorphism, $\varphi_t:\Sigma_t \rightarrow \Sigma$, from each spatial slice $\Sigma_t$ to the same model 3-manifold $\Sigma$ of constant curvature (i.e., to either the 3-sphere  $\mathbb{S}^3$, flat 3-space $\mathbb{R}^3$, or the hyperboloid $\mathbb{H}^3)$. This entails that the mapping $\varphi:M\rightarrow \mathbb{R} \times \Sigma, (t,p)\mapsto (t,\varphi_t(p))$ is a diffeomorphism. Otherwise stated, our spacetime manifold is globally diffeomorphic to the Cartesian product $ \mathbb{R} \times \Sigma$.

Returning to the classification of Finsler geometries in Section \ref{ssec:finsClass}, the situation is as follows:
\begin{itemize}
\item All homogeneous and isotropic pseudo-Riemannian geometries are given by the FLRW metric. In the language of this section, this means: $h(t,s) =\sqrt{ 1 - a(t)^2 s^2}$.
\item 
Homogeneous and isotropic Berwald geometries have been completely classified in \cite{Hohmann:2020mgs}. All non--pseudo-Riemannian Berwald structures are given by $h(t,s) = h(s)$ only. In particular, they are stationary, hence never describe a dynamically evolving universe.  Conversely,  all stationary cosmologically symmetric Finsler metrics are Berwaldian; this can be seen immediately by substituting $\partial_t h = \partial_t h' =0 $ in the general formula for the spray coefficients \eqref{eq:geod-hom1} and \eqref{eq:geod-hom2}.
\end{itemize}
For non-Berwald Landsberg geometries (the so-called \emph{unicorns}) the classification in homogeneous and isotropic symmetry is not known. This is what we will provide in the next section of this article.

\section{Solving the Landsberg condition}\label{sec:Unicorn}

In order to find all cosmologically symmetric (homogeneous and isotropic) Landsberg Finsler Lagrangians $L$, we will solve the equation given by the vanishing of the Landsberg tensor \eqref{eq:lands} in general
\begin{align}\label{eq:Lands0}
    P_{abc} = \nabla C_{abc} = 0\,.
\end{align}
To do this, it is convenient to first compute explicitly the relevant Finsler geometric objects (Cartan tensor, spray, nonlinear connection, etc.) for a generic homogeneous and isotropic Finsler Lagrangian \eqref{eq:homisoL2}. This has been carried out in Appendix \ref{app:HomAndIsoGeom}. In what follows, we will frequently refer to this Appendix whenever we use a result derived there. For example, using \eqref{eq:geod-hom1} and \eqref{eq:geod-hom2}, the dynamical covariant derivative of a function $h=h(t,s)$ is given by
\begin{align}
    \nabla h 
    &= \dot x^a \delta_a h 
    = \dot x^a \partial_a h - 2 G^a \dot \partial_a h\nonumber\\
    &= \dot t \partial_t h + \dot x^\alpha  \partial_\alpha h - 2 G^0 \dot \partial_0 h - 2 G^\alpha \dot \partial_\alpha h\nonumber\\
    &= \dot t \left( \partial_t h - \frac{h' \partial_t h'}{h''} \right)\,,
\end{align}
where we recall that $a,b =0,1,2,3$, whereas Greek letters $\alpha,\beta,..$ denote spatial indices.
The strategy to solve \eqref{eq:Lands0} is that we first identify some necessary conditions in section \ref{ssec:LandsCond}, which then give rise to six branches as candidates for solutions. In section \ref{ssec:solvLands} we solve each branch and verify that only one of these branches leads to three families of non-Berwald solutions of \eqref{eq:Lands0}.

\subsection{Deriving the necessary and sufficient Landsberg conditions}\label{ssec:LandsCond}
Taking into account that the time function $t$ is globally well defined (and so is the spatial metric $w$), we can construct the following well-defined tensor fields from the ``spatial metric" $w$  introduced in \eqref{eq:homisoL2}:
\begin{align}
	w_{\alpha\beta} = \frac{1}{2}\dot\partial_\alpha \dot\partial_\beta w^2\,,\quad
	\dot{\partial}_{\alpha}
        w = w_\alpha = \frac{1}{w} w_{\alpha\beta}\dot x^\beta\,,\quad
	w_a = (-s, w_\alpha)\,,\quad T= \dot t w\,.
\end{align}
These expressions make it convenient to display the Cartan tensor of a homogeneous and isotropic Finsler Lagrangian as a sum of two tensor fields (see \ref{sapp:HomIsoCartan} for the proof):
\begin{align}
    2 C_{abc} &=  \left( h (h' -  s h'') - s h'^2\right) T_{abc} + \frac{1}{\dot t} \left( h h''' + 3 h' h''\right) w_a w_b w_c\,,
\end{align}
where primes denote derivatives with respect to $s$ and $T_{abc} = \dot\partial_a \dot\partial_b \dot\partial_c T$. Next, the Landsberg tensor is easily obtained in terms of the dynamical covariant derivative using \eqref{eq:lands} and the relations
$\nabla w_a = p(t,s) \dot t w_a$ (the precise expression of $p(t,s)$, together with the proof of these equalities, are  also presented in the Appendix, in \eqref{eq:nablaw}), we find:
\begin{align}\label{eq:landshomiso}
    4 P_{abc} = \nabla \left( \left( h (h' -  s h'') - s h'^2\right) T_{abc} \right) + \left(\nabla \left(  \tfrac{( h h''' + 3 h' h'')}{\dot t}\right) + 3 \left( h h''' + 3 h' h''\right)p(t,s)   \right)w_a w_b w_c\,.
\end{align}
Since $T_{00c}=\dot\partial_0 \dot\partial_0 \dot\partial_c T =0$, the equations $P_{000} = 0$ and $P_{00\alpha} = 0$ are equivalent to each other and give a first necessary condition on $h$ in order for $L$ to be Landsberg:
\begin{align}\label{eq:nec1}
\boxed{    \nabla \left( \tfrac{( h h''' + 3 h' h'')}{\dot t}\right) = - 3 \left( h h''' + 3 h' h''\right)p(t,s)\,.
}
\end{align}
Consequently, the second term in \eqref{eq:landshomiso} vanishes and the remaining equations to be solved are:
\begin{align}\label{eq:uni2}
     \nabla (T_{abc}    \left( h (h' -  s h'') - s h'^2\right)) = 0\,.
\end{align}

The dynamical covariant derivative of $T_{abc}$ has the remarkable property that it is of the form: $\nabla T_{abc} = \dot t q(t,s) T_{abc}$ (see \eqref{eq:nablaT} for the expression of $q(t,s)$), meaning that the remaining part of the Landsberg condition is equivalent to:
\begin{align}
    T_{abc}\left[  \dot t q(t,s) \left( h (h' -  s h'') - s h'^2\right) + \nabla \left( h (h' -  s h'') - s h'^2\right) \right] = 0\,,
\end{align}
leading directly to the second necessary condition on $h$ in order for $L$ to be Landsberg:
\begin{align}\label{eq:nec2}
\boxed{    \nabla \left( h (h' -  s h'') - s h'^2\right) = - \dot t \left( h (h' -  s h'') - s h'^2\right) q(t,s)\,.}
\end{align}
Together, the two above conditions are also sufficient for the vanishing of $P_{abc}$. Briefly, the following equivalence holds: $P_{abc}=0$ if and only if \eqref{eq:nec1} and \eqref{eq:nec2} hold.

Now there are two cases to be distinguished, since the term
\begin{align}
    D=h s (- s h'^2 + h \bigl(h' + s h''))\partial_t h'
\end{align} will appear as a denominator.

\begin{itemize}
    \item Case I: $D=0$, leads to two branches of candidates for being Landsberg geometries
    \begin{align}
    0&=\partial_t h'  \quad &\Leftrightarrow \quad  h &= \alpha(t) + \beta(s)\,, \tag{Branch 1}\label{eq:denom1}\\
    0&=- s h'^2 + h \bigl(h' + s h''\bigr) \quad &\Leftrightarrow \quad h &=s^{c_1(t)}c_2(t)\,. \tag{Branch 2} \label{eq:denom2}
    \end{align}
    \item Case II: $D\neq0$, will lead to four branches of candidates for being Landsberg geometries. In this case equation \eqref{eq:nec2} can be algebraically solved for $h'''(t,s) = Q(t,s)$ as a function of the lower order derivatives of $h$, and \eqref{eq:nec1} can be algebraically solved for $h''''(t,s) = P(t,s)$ also as a function of the lower order derivatives. Thus, a consistency (integrability) condition for the system \eqref{eq:nec1} and \eqref{eq:nec2} is given in this case by $Q' = P$. Evaluating this consistency condition with Wolfram Mathematica yields
    \begin{align}\label{eq:integ}
    0&=\frac{h'' \left[s h'^2 + h \bigl(- h' + s h''\bigr)\right]\left[s h'' \partial_t h + \bigl(h -  s h'\bigr) \partial_t h'\right]}{h s \Bigl(- s h'^2 + h \bigl(h' + s h''\bigr)\Bigr) \partial_t h'}\\
    &\times\left[h'' h' \Bigl(s h'^2 + h \bigl(h' -  s h''\bigr)\Bigr) \partial_t h + 2 s h'' h  \Bigl(- h'^2 + h h''\Bigr) \partial_t h' -  h h' \Bigl(- s h'^2 + h \bigl(h' + s h''\bigr)\Bigr) \partial_t h''\right]\,.
    \end{align}
   This leads to the remaining branches, deemed Branch 3 - Branch 6. Branch 3 and 4 can be solved immediately:
    \begin{align}
        0&= h'' \quad &\Leftrightarrow \quad h &= c_1(t) + c_2(t) s\,, \tag{Branch 3}\label{eq:deg}\\
        0&=s h'^2 + h \bigl(- h' + s h''\bigr) \quad &\Leftrightarrow \quad h &= c_2(t)\sqrt{s^2 + c_1(t)}\,,\tag{Branch 4}\label{eq:pseudoRiem}
    \end{align}
    while Branch 5
    \begin{align}
        0&=s h'' \partial_t h + \bigl(h -  s h'\bigr) \partial_t h' \tag{Branch 5}\,,\label{eq:uni3}
    \end{align}
    and Branch 6
    \begin{align}
        0&=h' h'' \Bigl(s h'^2 + h \bigl(h' -  s h''\bigr)\Bigr) \partial_t h + 2 s h'' h  \Bigl(- h'^2 + h h''\Bigr) \partial_t h' \nonumber\\
        &-  h h' \Bigl(- s h'^2 + h \bigl(h' + s h''\bigr)\Bigr) \partial_t h''\tag{Branch 6}\label{eq:uni4}\,,
    \end{align}
    need further investigation.
\end{itemize}
The Finsler Lagrangian generated by \eqref{eq:deg} is degenerate, so not viable, while the one generated by \eqref{eq:pseudoRiem} leads to a pseudo-Riemannian geometry (which is trivially Landsberg and Berwald). We thus focus on the remaining four branches \eqref{eq:denom1}, \eqref{eq:denom2}, \eqref{eq:uni3} and \eqref{eq:uni4}. These require a more detailed discussion, which follows in the next section.

\subsection{Solving the necessary and sufficient Landsberg conditions}\label{ssec:solvLands}
The remaining task is to solve \eqref{eq:nec1} and \eqref{eq:nec2} for the four branches \eqref{eq:denom1}, \eqref{eq:denom2}, \eqref{eq:uni3} and \eqref{eq:uni4} of the integrabillity condition \eqref{eq:integ}.

\subsubsection{Three branches without unicorns}
As we will show below, the first three of the remaining branches \eqref{eq:denom1}, \eqref{eq:denom2} and \eqref{eq:uni3} all lead to either a degenerate fundamental tensor, or a non-Riemannian Berwald, thus stationary, geometry. Therefore, these branches do not lead to any candidates for a physically interesting model for the evolution of the universe.

\begin{itemize}
    \item \textbf{Branch 1,} $h = \alpha(t) + \beta(s)$: For this specific form of $h$, we find
    \begin{align}
        P_{000} = \frac{s^3 \dot \alpha}{4 (\alpha + \beta)}\left(3 \beta' \beta '' - (\alpha +\beta)\beta'''\right) = 0\,.
    \end{align}
    This term only vanishes if either $\dot \alpha = 0$, which implies that $L$ itself is independent of $t$ (and thus of Berwald type, see \cite{Hohmann:2020mgs}), or if $\left(3 \beta' \beta '' - (\alpha +\beta)\beta'''\right)=0$. Differentiating the latter with respect to $t$ implies $\dot{\alpha} \beta''' =0$. Since, as already noted, $\dot{\alpha}=0$ does not lead to any convenient (non-Berwald) solution, the only nontrivial remaining possibility is that $\beta''' =0$. When substituted back into the given equation an immediate consequence is $ \beta' \beta '' =0$. In turn, this implies that $\beta = c_1 + c_2s$. In other words we find $h(t,s) = \alpha(t) + c_1 + c_2s$, being of the same form as \eqref{eq:deg} which yields a degenerate Finsler Lagrangian.

    \item \textbf{Branch 2,} $h=s^{c_1(t)}c_2(t)$:  Evaluating the Landsberg tensor for the second branch \eqref{eq:denom2}, the Landsberg tensor vanishes only if $c_1(t) = c_1$ is constant. We thus obtain the Finsler Lagrangian
    \begin{align}\label{eq:uni1}
        L = \dot t^2 s^{c_1} c_2(t)\,,
    \end{align}
   which is of Berwald type, since it is of the form identified in \cite{Hohmann:2020mgs} (and the $t$-dependence can be eliminated by a convenient redefinition of the $t$ coordinate). Moreover, it never defines a Finsler spacetime structure, since $L=0$ does not lead to well-defined convex causal cones.

    \item  \textbf{Branch 5},  $ s h'' \partial_t h + (h -  s h') \partial_t h'=0$: Let us first assume that $h-sh' \neq 0$, which allows us to rewrite the above equation as:
    \begin{align}
        \partial_s\left( \frac{\partial_t h}{h - s h'} \right) = 0\,,
    \end{align}
    which leads to $c(t) (h-sh') = \partial_t h$.
    Defining $q = h/s$,  we obtain
    \begin{align}
        \partial_s q
        = \partial_s \left(\frac{h}{s}\right) = \frac{s h' - h}{s^2} = - \frac{\partial_t h}{s^2 c(t)}
        = - \frac{1}{s c(t)} \partial_t q\,,
    \end{align}
    that is: $s \partial_s q + c(t)^{-1} \partial_t q = 0$. The latter can be solved by the methods of characteristics giving
    \begin{align}
        q(t,s) = q(u(t,s))\,,\quad u(t,s) = s e^{-\int_0^t c(t') dt'}\,,
    \end{align}
    and thus
    \begin{align}
        h(t,s) = q(u(t,s)) s \,.
    \end{align}
    Unfortunately, the corresponding Finsler Lagrangian,
    \begin{align}
        L = \dot t^2\ q(u(t,s))^2\ s^2\,,\quad u(t,s) = s e^{-\int_0^t c(t') dt'}\,,
    \end{align}
    is not only Landsberg, but of Berwald type, since the apparent $t$-dependence through the factor $e(t) := e^{-\int_0^t c(t') dt'}$ can be absorbed by a redefinition of the $t$-coordinate as $\dot{\eta}:= \dot t / e(t)$, and then the results from \cite{Hohmann:2020mgs} apply.
    The case when $h-sh'=0$ leads to: $h=c_1 (t)s$, which is degenerate.
\end{itemize}

Isolated singularities can occur either on spacetime or along tangent directions. For spacetime singularities, e.g. , $c(t)=0$, we can specify a physical meaning (Big Bang), whereas for singular directions, this is still to be investigated.

\subsubsection{Branch 6: The unicorn branch}
Having analyzed the above branches, we realize that $4$-dimensional homogeneous and isotropic unicorns (if any) must be given by \eqref{eq:uni4}. In order to solve the Landsberg condition for this branch, we employ the following strategy. First we solve \eqref{eq:uni4} for $\partial_t h''$, which gives:
\begin{align}\label{eq:dth1}
    \partial_t h'' = \frac{h'' \biggl(h' \Bigl(s h'^2 + h \bigl(h' -  s h''\bigr)\Bigr) \partial_t h + 2 h s \Bigl(- h'^2 + h h''\Bigr) \partial_t h'\biggr)}{h h' \Bigl(- s h'^2 + h \bigl(h' + s h''\bigr)\Bigr)}\,.
\end{align}
Second, we also solve \eqref{eq:nec2} for $\partial_t h''$, and get as alternative expression
\begin{align}\label{eq:dth2}
    \partial_t h'' = \frac{h''\Bigl(s h'^2 + h \bigl(h' -  s h''\bigr)\Bigr)\partial_t h}{h\Bigl(- s h'^2 + h \bigl(h' + s h''\bigr)\Bigr)}+ \left(\frac{h'\Bigl(2 h^2 -  s^2 h'^2 + h s \bigl(- h' + s h''\bigr)\Bigr)}{sh \Bigl(- s h'^2 + h \bigl(h' + s h''\bigr)\Bigr)}+ \frac{h'''}{h''}\right)\partial_t h'.
\end{align}
Equating these two necessary expressions for $\partial_t h''$ gives a simplified necessary condition, noting that the terms proportional to $\partial_t h$ cancel, and the other terms combine to
\begin{align}\label{eq:nec3}
    \left(\frac{2}{s} + \frac{h'}{h} -  \frac{2 h''}{h'} + \frac{h'''}{h''}\right) \partial_t h' = 0\,.
\end{align}
The case when $\partial_t h' = 0$ on some open subset of the domain of definition of $h$ was already discussed in \eqref{eq:denom1} and it leads (assuming nondegeneracy) to a Berwald metric. Physically speaking, this would correspond to the existence of a region of spacetime where the Finsler metric is stationary. Excluding such scenarios, we may assume that $\partial_t h' \neq 0$ and equate the bracket in \eqref{eq:nec3} to zero,
\begin{align}
    \frac{2}{s} + \frac{h'}{h} -  \frac{2 h''}{h'} + \frac{h'''}{h''}= 0\,.
\end{align}

A first integral is,
\begin{align}
        c_1(t) = \frac{s^2 h h''}{h'^2} \Leftrightarrow \partial_s\left(\frac{h}{h'}\right) = 1 - \frac{c_1(t)}{s^2}\,,
\end{align}
which can in turn be integrated to
\begin{align}\label{eq:h4}
    \ln(h) &= c_3(t)+\int \frac{s}{s^2 + c_2(t) s + c_1(t)} ds \\
    \Leftrightarrow h &= e^{c_3(t)} \exp\left[\int \frac{s}{s^2 + c_2(t) s + c_1(t)} ds\right] \,.
\end{align}
In order to evaluate the integral, we distinguish three cases:
\begin{enumerate}
    \item For $c_1(t)> c_2(t)^2/4$, the integral can be solved to be
    \begin{align} \label{eq:L1}
    h = e^{c_3(t)} \sqrt{s^2 + c_2(t) s + c_1(t)}
        \exp\left[- \frac{c_2(t)}{\sqrt{4 c_1(t) - c_2(t)^2}} \arctan\left( \frac{2 s + c_2(t)}{\sqrt{4 c_1(t) - c_2(t)^2}} \right) \right]\,.
    \end{align}
    Using this expression in equation \eqref{eq:nec1} we find that we obtain a non-Berwald Landsberg Finsler geometry if and only if $c_1(t) = d_1 c_2(t)^2$, where $d_1 > 1/4$ is a constant. Thus, with such a choice of $d_1$,
    \begin{align}\label{eq:LU1}
        L_1 = \dot t^2 e^{2 c_3(t)}(d_1 c_2(t)^2 + s c_2(t) + s^2)
        \exp\left[- \frac{2 }{\sqrt{4 d_1 - 1}} \arctan\left( \frac{2 s + c_2(t)}{\sqrt{4 d_1 - 1}c_2(t)} \right) \right]
    \end{align}
    defines a unicorn Finsler geometry. Defining a new conformal time coordinate $\eta$ via $\dot{\eta} =d_2 c_2(t) \dot t$ (and using $\tilde s = w/ \dot{\eta}$) transforms the unicorn to a conformal separated variables form
    \begin{align}\label{eq:LU1-2}
        L_1 = \dot{\eta}^2 \Omega_1(\eta)^2 f_1(\tilde s)
        = \dot{\eta}^2 e^{2 c_3(t(\eta))} \left(\frac{d_1}{d_2^2} + \frac{1}{d_2} \tilde s + \tilde s^2\right)
        \exp\left[- \frac{2 }{\sqrt{4 d_1 - 1}} \arctan\left( \frac{2 \tilde s d_2+ 1}{\sqrt{4 d_1 - 1}} \right) \right]\,.
    \end{align}

    \item For $c_1(t)< c_2(t)^2/4$, the integral is solved analogously to the previous case, just changing $\arctan$ to $\text{arctanh}$:
        \begin{align}
        h &= e^{c_3(t)} \sqrt{s^2 + c_2(t) s + c_1(t)}
        \exp\left[- \frac{c_2(t)}{\sqrt{c_2(t)^2- 4 c_1(t)}} \text{arctanh}\left( \frac{2 s + c_2(t)}{\sqrt{c_2(t)^2-4 c_1(t)}} \right) \right]\\
        &= e^{c_3(t)} (s_1(t) - s)^{\frac{s_1(t)}{s_1(t)-s_2(t)}} (s - s_2(t))^{\frac{s_2(t)}{s_2(t)-s_1(t)}}  \,.
        \end{align}
       To get the second equality above, we used the relation $\exp(d_1\ {\rm arctanh}(X)) = ((1+X)/(1-X))^{d_1/2}$ and introduced
        \begin{align}
            s_1(t) s_2(t) = c_1(t),\quad s_1(t) + s_2(t) = - c_2(t)\,.
        \end{align}
        Evaluating again \eqref{eq:nec1} for this solution of the necessary conditions, we find that we obtain a Landsberg Finsler geometry if and only if $s_1(t) = d_1 s_2(t)$, with $d_1 \in \mathbb{R}$. If $d_1=-1$ , $L$ is  pseudo-Riemannian - i.e., it is the Finsler function of the FLRW metric. Yet,  for $d_1\neq -1$, we obtain the Finsler Lagrangian:
        \begin{align}\label{eq:LU2}
            L_2 = \dot t^2 e^{2 c_3(t)} (d_1 s_2(t) - s)^{\frac{2 d_1}{d_1-1}} (s - s_2(t))^{\frac{2 }{1-d_1}}
        \end{align}
       Again, a redefinition of the time coordinate by $\dot{\eta} = \dot t d_2 s_2(t)$ to conformal time $\eta$, where $d_2=const.$, casts it into a separated variables form:
        \begin{align}\label{eq:LU2-2}
            L_2 = \dot{\eta}^2 \Omega_2(\eta)^2 f_2(\tilde s)
            = \dot {\eta}^2 e^{2 c_3(t(\eta))} (d_1 - \tilde s d_2)^{\frac{2 d_1}{d_1-1}} (\tilde s d_2 - 1)^{\frac{2 }{1-d_1}} \frac{1}{d_2^2}\,.
        \end{align}
        From this expression one nicely sees that it is non-Berwaldian (see the spray coefficients below in \eqref{eq:sprayUni} with $e_1 = d_1/d_2$ and $e_2 = 1/{d_2}$), in other words, it defines a unicorn Finsler geometry.
        \item For $c_1(t) = c_2(t)^2/4$ the integration in \eqref{eq:h4} yields
        \begin{align}
            h = e^{c_3(t)}(2 s + c_2(t))\exp\left( \frac{c_2(t)}{2 s + c_2(t)}\right)\,.
        \end{align}
        Evaluating again \eqref{eq:nec1} for this solution of the necessary conditions, we find immediately that we obtain a non-Berwald Landsberg-Finsler geometry. Thus,
        \begin{align}\label{eq:LU3}
            L_3 = \dot t^2 e^{2 c_3(t)} (2 s + c_2(t))^2\exp\left( \frac{ 2 c_2(t)}{2 s + c_2(t)}\right)
        \end{align}
        defines a unicorn Finsler geometry. As for the other two unicorn classes, we can redefine the time coordinate to conformal time $\eta$ by: $\dot {\eta} = \dot t c_2(t) d_2$ (where $d_2=const.$), to express the Lagrangian as
        \begin{align}\label{eq:LU3-2}
            L_3 = \dot{\eta}^2 \Omega_3(\eta)^2 f_3(\tilde s)
            =\dot{\eta}^2 e^{2 c_3(t(\eta))} (2 \tilde s d_2 + 1)^2\exp\left( \frac{ 2 }{2 \tilde s d_2+ 1}\right)\frac{1}{d_2^2}\,.
        \end{align}
\end{enumerate}
Having analyzed the Landsberg tensor for general $4$-dimensional homogeneous and isotropic Finsler Lagrangians, we have found three classes of unicorn geometries given by the Finsler Lagrangians $L_1$ \eqref{eq:LU1}, $L_2$ \eqref{eq:LU2} and $L_3$ \eqref{eq:LU3}. \vspace{11pt}

Since we solved the equation $P_{abc} = 0$ in all generality, there cannot exist any more classes. 
In other words, we have found \emph{all} $4$-dimensional homogeneous and isotropic unicorns. \vspace{11pt}

\noindent In earlier works, a different strategy was used to find unicorns. Starting from a Berwald Finsler Lagrangian $L_0$, unicorn Finsler geometries were constructed by a conformal rescaling
\begin{align}\label{eq:confberw}
    L_U = e^{\sigma(x)} L_0\,,
\end{align}
and solving the so-called $\sigma T$ condition \cite{elgendi_2020,Elgendi2021a,Elgendi2021b}. This way, several unicorns were found. Due to the ansatz $L_U = e^{\sigma(x)} L_0$ this method can, yet, not ensure that all possible unicorns of a certain type are found, since it assumes the specific structure \eqref{eq:confberw}. In the 4-dimensional homogeneous and isotropic scenario we study here, this ansatz would amount to assume a separated variable Finsler Lagrangian from the beginning
\begin{align}
    L_U = e^{\sigma(t)} L_0(s)\,.
\end{align}
Solving the $\sigma T$ condition for this form of the Finsler Lagrangian leads to the same unicorn geometries $L_1$, $L_2$ and $L_3$ that we found. We note that $L_1$ had already been found earlier in \cite{shen_unicorns} and in \cite{ASANOV2006275}. In addition, our approach proves that there cannot be others.

Hence our results completely classify and present explicitly all $4$-dimensional homogeneous and isotropic unicorn geometries.

Next we determine which of these classes of unicorns really define a Finsler spacetime.

\section{Non-Berwald Landsberg  Finsler spacetimes}\label{sec:UnicornSpacetime}
Having identified all homogeneous and isotropic non-Berwald Landsberg Finsler Lagrangians, we are interested in the question which ones lead to a viable Finsler spacetime, i.e.\ which ones can be used as spacetime geometry in cosmology? The answer to this question leads us to the unique Finsler non-Berwald Landsberg (or Finsler unicorn) generalization of FLRW geometry.

In order to verify which Finsler Lagrangian satisfies the Finsler spacetime definition discussed in the beginning of Section \ref{ssec:FGeom}, 
we determine their null structure $L=0$ and verify the Lorentzian signature of the Finsler metric by the sign of its determinant
\begin{align}
    \det g = \det w_{\alpha\beta} \frac{h^4 h'^2 }{s^2}\ h h''\,.
\end{align}

Let us begin with $L_1$ and $L_3$, for which we immediately see that they do not yield any realistic light cones, hence no Finsler spacetime structure. Indeed:
\begin{itemize}
    \item The Finsler Lagrangian \eqref{eq:L1} can never vanish, since 
    \begin{align}
        s^2 + c_2(t)s + c_1(t) \neq 0\,,
    \end{align}
    for the required constraints on the parameters $c_1(t)$ and $c_2(t)$. Thus, $L_1$ never defines a Finsler spacetime.

    \item The Finsler Lagrangian of the type \eqref{eq:LU3-2} can be expressed as
    \begin{align}
        L_3(\eta,\tilde s)
        =\dot{\eta}^2 e^{2 c_3(\eta)} (2 \tilde s d_2 + 1)^2\exp\left( \frac{ 2 }{2  \tilde s d_2+ 1}\right)\frac{1}{d_2^2}\,.
    \end{align}
    A direct calculation of the determinant of the Finsler metric shows that it is positive. Thus, $L_3$ does not lead to a Finsler spacetime structure\footnote{In \cite{Heefer:2023a} possible modifications of Finsler Lagrangians of the type $L_3$ have been discussed as solutions of the Finsler gravity equation.}.
\end{itemize}

From the perspective of a physical spacetime structure, the only possibility remains thus $L_2$ \eqref{eq:LU2-2}, which, introducing the constants $e_1 = d_1/d_2$ and $e_2 = 1/{d_2}$ with $e_1\neq e_2$, can be expressed as
\begin{align}\label{eq:LU2-3}
    L_2(\eta,\tilde s) =
    \dot {\eta}^2 e^{2 c_3(\eta)} ( e_1 -  \tilde s )^{\frac{2 e_1}{e_1-e_2}} ( \tilde s - e_2)^{-\frac{2 e_2 }{e_1-e_2}} \,.
\end{align}
The  lightlike directions of this geometry are given by $\tilde s=e_1$ and $\tilde s=e_2$. In order to have both a future and a past pointing lightcone, we need that $e_1 e_2 <0$. Moreover, we need that the exponents $\frac{2 e_1}{e_1-e_2}$ and $-\frac{2 e_2 }{e_1-e_2}$ are both positive. This is ensured by either choosing $e_1<0, e_2>0$ and $e_1-e_2<0$, or choosing $e_1>0, e_2<0$ and $e_1-e_2>0$. Both choices lead to the same results and thus the geometry defined by $L_2$ has a well defined light cone structure. It is clear that $L_2$ does not smoothly extend to the light cones except $e_1=-e_2$, that is $L_2$ becomes the Finsler Lagrangian of classical FLRW geometry. However, there do exist choices of $e_1$ and $e_2$ (for example $e_1, e_2 \in \mathbb{Z}$) such that a power of $L_2$, $\tilde L_2 = L_2^{(e_1-e_2)}$, is well defined and smooth on the lightcone. It has been shown in \cite{Pfeifer:2011tk}, that $\tilde L_2$ and $L_2$ give the same canonical non-linear connection coefficients, which are smooth on the lightcone. Thus, in this sense, the geometry of $(M,L_2)$ is well defined on the light cones.

Next, studying the determinant of the Finsler metric for this Finsler Lagrangian, identifying $h(\eta,\tilde s) = e^{c_3(\eta)} ( e_1 -  \tilde s )^{\frac{ e_1}{e_1-e_2}} ( \tilde s - e_2)^{-\frac{ e_2 }{e_1-e_2}}$ from \eqref{eq:LU2-3}, yields
\begin{align}
    \det g
    &= \det w_{\alpha\beta}\ \frac{h^4 h'^2 }{\tilde s^2}\ h h''\nonumber\\
    &=\det w_{\alpha\beta}\  e^{2c_3(\eta)} \ (e_1 e_2 )\ (e_1-\tilde s)^{4\frac{(e_1+e_2)}{e_1-e_2}} (\tilde s-e_2)^{-4\frac{e_1+e_2}{e_1-e_2}}\,.
\end{align}
This determinant is clearly negative, since the sign is determined by the factor $e_1 e_2$, and smooth until it reaches the light cones. Thus, the Finsler metric has Lorentzian signature in the interior of both, the future and the past pointing light cone. Depending on the exact values of $e_1$ or $e_2$, it vanishes at one of the light cones $s=e_1$ and blows up at the other $s=e_2$.

In order to prove that signature of the metric tensor $g$ of $L_2$ is $(+,-,-,-)$ inside the timelike cones, we note that all cosmologically symmetric Finsler metric are generalized $(\alpha,\beta)$-metrics, i.e.\ Finsler Lagrangians that are functions $L=L(b,\alpha,\beta)$, where $\alpha = e^{c_3(\eta)} \sqrt{|\dot{\eta}^2 - w^2|}$ is the FLRW metric (with signature convention $(+,-,-,-)$ written in conformal time), $\beta = \dot{\eta}$ and $b=a_{ij}b^ib^j =e^{-c_3(\eta)} $. Then, using Lemma 3 in \cite{Voicu:2023zem}, which still can be applied since $b$ does not depend on $\dot x$, it turns out that a necessary and sufficient condition for the signature condition of $L_2$ is that $det(g)<0$, which is satisfied as we just have proven above.

It is remarkable that even if the determinant is not well defined on one of the light cones, the geodesic equation, expressed in terms of the geodesic spray, is well defined. Thus, all lightlike geodesics can be described without any problems. To see this, we display below the geodesic spray explicitly, by evaluating the expression in the Appendix \ref{sapp:HomIsoGandN} for \eqref{eq:LU2-3},
\begin{align}\label{eq:sprayUni}
    G^0 &= \dot \eta^2 \partial_\eta c_3(\eta) \frac{(e_1 e_2 - \tilde s^2)}{e_1 e_2}\,,\\
    G^\alpha  &= \tilde G^\alpha + \dot \eta \dot x^\alpha \frac{\partial_\eta c_3(\eta)}{2}  \frac{(2 e_1 e_2 - (e_1+e_2)\tilde s)}{2 e_1 e_2}\,.
\end{align}
We clearly see that  $G^0$ and $G^{\alpha}$  are well defined for $\tilde s\to e_1$ or $\tilde s\to e_2$. The same holds for all timelike directions, in particular for $\tilde s = 0$, i.e.\ for geodesics $\gamma$ with tangents $\dot \gamma \sim \partial_\eta$ proportional to the conformal time direction.

As the Finsler Lagrangian is expressed in \eqref{eq:LU2-3}, the time coordinate can be interpreted as conformal cosmological time. Another redefinition of $\eta$ and the appearing constants defined by
\begin{align}
    \dot {t} = a(\eta) \dot \eta e_2, \quad a(\eta) = e^{c_3(\eta)},\quad e_1 = f e_2
\end{align}
enables us to rewrite the physical Finsler Lagrangian in terms of standard cosmological time $t$
\begin{align}\label{eq:FinslerUnicornFLRW}
\boxed{
    L_2
    =  \left( \dot {t} f -  a(t) w \right)^{\frac{2 f}{f-1}} \left( a(t) w - \dot {t} \right)^{-\frac{2  }{f-1}}\,.}
\end{align}

In conclusion, among all the 4-dimensional homogeneous and isotropic non-Berwald Landsberg (unicorn) geometries, we identified one which is physically viable.  It possesses one dynamical variable, a scale factor $a(t)$, and two independent constants: the spatial curvature parameter $k$, and a Finsler parameter $f$. For $f=-1$, the Finsler Lagrangian $L_2$ becomes pseudo-Riemannian FLRW geometry.

Moreover, our findings gives us new insights about the definitions of Finsler spacetimes. The Finsler manifold $(M,L_2)$ defined by \eqref{eq:FinslerUnicornFLRW} does not satisfy the definition of a Finsler spacetime given in Section \ref{ssec:FGeom} since it is not smooth on all timelike directions $\mathcal{T}_x$ (the $\dot x$-derivatives are not defined at $\dot r = \dot \theta = \dot \phi = 0$). This points towards the conclusion that the more general definition employed in \cite{Caponio-Stancarone}, which allows for certain non-smoothness of $L$ inside the cone of timelike directions, is the more appropriate physically useful definition of Finsler spacetimes. Nonetheless, this non-smoothness of $L_2$ does not cause any problems, since all timelike geodesics can be calculated, as we discussed above.

With this discussion, we identified and introduced the \emph{unique} Finsler non-Berwald Landsberg (or Finsler unicorn) generalisation of FLRW geometry.

\section{Discussion and conclusion}\label{sec:disc}
In order to prove that the neglected contributions of the 1PDF of a kinetic gas to the gravitational field of the kinetic gas can be the sources of dark matter and dark energy, the goal is to solve the Finsler gravity equation \eqref{eq:Fgrav} in homogeneous and isotropic symmetry. This endeavour turns out to be very involved, even in this highly symmetric situation, due to the complexity of the equations and the generally large number of degrees of freedom of a function $h(t,s)$ (instead of only a scale factor $a(t)$ in general relativity) in Finsler geometry. \vspace{11pt}

Our identification of the \emph{unique} Finsler non-Berwald Landsberg (or Finsler unicorn) generalisation of FLRW geometry \eqref{eq:FinslerUnicornFLRW} is a significant step towards finding cosmological solutions of the Finsler gravity equation. \vspace{11pt}

For non-Berwald Landsberg geometries, the second part of the field equation \eqref{eq:Fgrav} (which contains the trace of the Landsberg tensor $P_a$) is identically zero, leaving only the non-trivial curvature part. Moreover, the number of degrees of freedom that need to be determined by the equation is drastically reduced to just one time-dependent scale factor, $a(t)$, and a constant parameter $f$. We demonstrated this by determining all possible homogeneous and isotropic unicorns directly from the vanishing of the Landsberg tensor. To do so we used the geometric tensors of 4-dimensional homogeneous and isotropic Finsler geometry. Their complete expressions are derived explicitly for the first time and can be found in Appendix \ref{app:HomAndIsoGeom}. These will be extremely useful in the future for the derivation of the Finsler Friedmann equations, in general and for the Finsler extension of FLRW geometry presented here.

A detailed analysis of the physical consequences of the new Finslerian FLRW spacetime geometry will be presented in a follow-up article. One remarkable property is a difference in the opening angles of the future and past light cones, i.e., a time asymmetry in the spacetime geometry. This might be connected to the direction of the flow of time in the future and deserves thorough investigation.

From our finding that there is no cosmological non-pseudo-Riemannian Finsler spacetime unicorn geometry that is smooth on the lightcone, we are inspired to formulate the following mathematical conjecture, which is to be proven or disproven in the future. Conjecture: There exists no non-Berwald Landsberg (or Finsler unicorn) spacetime geometry, that is smooth on its light cones.

Last, but not least, the study of homogeneous and isotropic Finsler manifolds introduces a new aspect into the discussion on the optimal formulation of the definition of Finsler spacetimes. Since generically, for all homogeneous and isotropic Finsler manifolds, not all $\dot x$-derivatives of the Finsler Lagrangian $L$ are well defined at $\dot r = \dot \theta = \dot \phi = 0$ ($w=0$), the definition presented in \cite{Caponio-Stancarone} seems favorable over the more restrictive one that demands smoothness of $L$ on all timelike directions.

In conclusion, with this work we are one step closer to understand the details of the description of the gravitational field of kinetic gases in term of Finsler geometry instead of pseudo-Riemannian geometry. Our work lays the foundation for solving the Finsler gravity equation sourced by the 1PDF of a kinetic gas in cosmology, to describe the evolution of the universe. Moreover, we gave a complete mathematical classification of all 4-dimensional homogeneous and isotropic non-Berwald Landsberg Finsler geometries and opened the path to a deeper understanding of the properties of this class of Finsler geometries in general.


\begin{acknowledgments}
The authors would like to thank Andrea Fuster and Volker Perlick for useful discussions and insights.
CP acknowledges support by the excellence cluster QuantumFrontiers of the German Research Foundation (Deutsche Forschungsgemeinschaft, DFG) under Germany's Excellence Strategy -- EXC-2123 QuantumFrontiers -- 390837967 and was funded by the Deutsche Forschungsgemeinschaft (DFG, German Research Foundation) - Project Number 420243324 - and by the Transilvania Felowships for Visiting Professors grant 2024 of the Transilvania University of Brasov.
This work is based upon collaboration within the COST Action CA 21136, “Addressing observational tensions in cosmology with systematics and fundamental physics” (CosmoVerse).
\end{acknowledgments}

\appendix

\section{Homogeneous and isotropic Finsler geometry}\label{app:HomAndIsoGeom}
In this appendix, we explicitly list the geometric tensors of Finsler geometry in homogeneous and isotropic symmetry. They build the foundation on which all derivations in this article are based.

\subsection{Finsler metric and Cartan tensor}\label{sapp:HomIsoCartan}
In order to find the homogeneous and isotropic unicorns in Sec. \ref{sec:Unicorn}, we need the Finsler metric and the Cartan tensor for cosmologically symmetric Finsler Lagrangians $L$, which are all described by:
\begin{align}
	L = \dot t^2 h(t,s)^2\,, \quad s=\frac{w}{\dot t}\,,\quad w^2 = \frac{\dot r^2}{1-k r^2}+r^2\big(\dot \theta^2+\sin^2(\theta) \dot\phi^2\big)\,.
\end{align}
Introducing the notations,
\begin{align}
w_{\alpha\beta} = \frac{1}{2}\dot\partial_\alpha \dot\partial_\beta w^2
= \textrm{diag}\left(\frac{1}{1-kr^2},r^2,r^2 sin^2(\theta)\right)\,,
\quad
\dot{\partial}_{\alpha} w = w_\alpha = \frac{1}{w} w_{\alpha\beta}\dot x^\beta\,,
\quad
w_a = (-s, w_\alpha)\,,
\end{align}
which satsify the following relations
\begin{align}\label{eq:homw}
   \dot x^a w_a = -s \dot t + \dot x^\alpha w_\alpha = - w + w =0\,, 
\end{align}
we first obtain:
\begin{align}
	\dot{\partial}_0 s = -\frac{s}{\dot t}\,,\quad
	\dot{\partial}_\alpha s = \frac{w_\alpha}{ \dot t}\,.
\end{align}
Denoting by primes derivatives with respect to $s$, we find:
\begin{itemize}
    \item The components of the Finsler metric and its inverse:
    \begin{align}\label{metric_comps_cosmo}
	g_{00}&=h^{2}-2shh'+s^{2}(h'{}^{2}+h h'')\,,\\
	g_{0\alpha}&=w_{\alpha}\left(h h'-s(h'{}^{2}+h h'')\right)\,,\\
	g_{\alpha\beta}&=h h' \frac{w_{\alpha\beta}}{s}+\left(s h'{}^{2}+s h h''-h h'\right)\frac{w_{\alpha}w_{\beta}}{s}\,,
    \end{align}
    and
    \begin{align} \label{eq:inverse_metric}
	g^{00}&=A\,, &  A&=\frac{h'^2+ h h''}{h^3 h''}\,, \nonumber\\
	g^{0\alpha }&=B\frac{\dot{x}^{\alpha }}{w}\,, & B&=\frac{s (h'^2+ h h'') - h h'}{h^3 h''}\,,\\
	g^{\alpha \beta }&=C w^{\alpha \beta }+\frac{D}{w^{2}}\dot{x}^{\alpha }\dot{x}^{\beta }\,, & C&=\frac{s}{ hh{' }}\,, D=\frac{(h-sh')\left(hh'-s(h'{}^{2}+hh'')\right)}{h^{3}h'h''}\,,\nonumber
    \end{align}
    where $w^{\alpha\beta}=\text{diag}(1-kr^2,\frac{1}{r^2},\frac{1}{r^2sin^2(\theta)})$  is the inverse of $w_{\alpha\beta}$.
    \item The Cartan tensor:
    \begin{align}
	C_{000} &= -\frac{1}{2} \frac{s^3}{\dot t} ( h h''' + 3 h' h'')\,,\\
	C_{00\alpha} &= w_\alpha \frac{s^3}{2 w} ( h h''' + 3 h' h'')\,,\\
	C_{0\alpha\beta}
        &= \frac{w_{\alpha\beta}}{2 w}\left( h (h'-s h'')-  s h'^2 \right) - \frac{w_\alpha w_\beta}{2 w}\left( h(h'-sh'')- s h'^2 + s^2 (h h''' + 3 h' h'') \right)\,,\\
	C_{\alpha\beta\gamma} &= - \frac{(w_{\beta\gamma} w_{\alpha} + w_{\alpha\gamma} w_{\beta} + w_{\alpha\beta}        w_{\gamma} - 3 w_{\alpha} w_{\beta} w_{\gamma}) \left( h (h' -  s h'') - s h'^2\right)}{2 s w} \nonumber\\
	&+ \frac{s w_{\alpha} w_{\beta} w_{\gamma} ( h h''' + 3 h' h'')}{2 w} \,.
    \end{align}
    Introducing the scalar variable $T = \dot t w$, this can be combined to a full Finsler spacetime tensor as
    \begin{align}
        C_{abc}
        &= \frac{1}{2} \left( h (h' -  s h'') - s h'^2\right) T_{abc} + \frac{\left( h h''' + 3 h' h''\right)s}{2 w} w_a w_b w_c\,,
    \end{align}
    where:
    \begin{align}
        T_{abc} &:= \dot\partial_a \dot\partial_b \dot\partial_c T\\
        &= (\delta_c^0 \delta_b^\beta \delta_a^\alpha  + \delta_a^0 \delta_b^\beta \delta_c^\alpha +\delta_b^0 \delta_c^\beta \delta_a^\alpha) \left(\tfrac{w_{\alpha\beta}-w_\alpha w_\beta}{w}\right)\nonumber\\
        &+ \frac{\dot t}{w^2}\delta_c^\gamma \delta_b^\beta \delta_a^\alpha (3 w_\alpha w_\beta w_\gamma - w_{\alpha\beta} w_\gamma - w_{\alpha\gamma} w_\beta - w_{\gamma\beta} w_\alpha)\,.
    \end{align}
\end{itemize}

\subsection{Geodesic spray components and nonlinear connection coefficients}\label{sapp:HomIsoGandN}
In order to derive the Landsberg tensor in equation \eqref{eq:landshomiso}, we need to evaluate dynamical covariant derivatives, which employ geodesic spray coefficients and nonlinear connection coefficients \eqref{eq:geodspray}. Here we display their explicit form for homogeneous and isotropic Finsler Lagrangians.

From the definition of the geodesic spray \eqref{eq:geodspray} we find:
\begin{align}\label{eq:geod-hom1}
    G^0 = \dot t^2 \frac{\bigl(h'' \partial_t h -  h' \partial_t h'\bigr)}{2 h h''}
\end{align}
and
\begin{align}\label{eq:geod-hom2}
    G^\alpha =   \tilde G^\alpha
    + \frac{1}{2} \dot t \dot{x}^{\alpha}
    \frac{ \Bigl(s h'' \partial_t h + \bigl(h -  s h'\bigr) \partial_t h'\Bigr) }{s h  h''}\,,
\end{align}
where the quantities $\tilde G^\alpha$ are the geodesic spray coefficients of the 3-dimensional Finsler function $w^2$, given by
\begin{align}
	\tilde G^\alpha = \frac{1}{4}w^{\alpha\beta}\left( \dot x^\gamma \dot{\partial}_\beta \partial_{\gamma} w^2- \partial_\beta w^2\right)\,.
\end{align}
The nonlinear connection coefficients are then obtained by $N^a{}_b = \dot \partial_b G^a$, as follows:
\begin{align}\label{eq:nonlin-hom}
    N^0{}_0
    &= \frac{\dot t}{2 h^2 h''^2}\left[ h''^2(sh'+2h)\partial_t h - h'\bigl((sh'+2)h''+shh'''\bigr) \partial_t h'+sh'h''\partial_t h''\right]\,,\\
    N^0{}_\alpha
    &= \frac{\dot t w_{\alpha} h'}{2 s h^2 h''^2}  \left[ -h''^2\partial_t h + (h'h''+hh''')\partial_t h'-hh''\partial_t h''\right]\,,\\
    N^\alpha{}_0
    &= \frac{\dot{x}^{\alpha}}{2 h^2 h''^2}
    \biggl[ sh''^2(h+sh')\partial_t h + \Bigl( h''\bigl( 2h^2-sh'(h+sh')\bigr)+shh'''(h-sh')\Bigr)\partial_t h'\nonumber\\
    &- sh^2h''(h-sh')\partial_t h''\biggr],\\
    N^\alpha{}_\beta
    &= \Gamma^{\alpha}{}_{\beta\gamma} \dot{x}^{\gamma}
    + \frac{\dot t \delta^{\alpha}_{\beta}}{2 s h h''} \left[s h'' \partial_t h + \bigl(h -  s h'\bigr) \partial_t h'\right] - \frac{w_{\beta} \dot{x}^{\alpha}}{2 s^2  h''^2} \left[(h''+sh''')\partial_t h'-s\partial_t h'' \right]  \nonumber\\
    &+ \frac{w_{\beta} \dot{x}^{\alpha} h'}{2 h^2  h''^2} \left[-h''^2\partial_t h + (h'h''+hh''')\partial_t h'-hh''\partial_t h''\right],
\end{align}

where $\Gamma^{\alpha}{}_{\beta\gamma}$ are the coefficients of the Levi-Civita connection of the spatial metric with components $w_{\alpha\beta}$.

By means of these expressions, one can directly determine the following terms, which we used to derive the Landsberg tensor \eqref{eq:landshomiso} and study the necessary equations \eqref{eq:nec1} and \eqref{eq:nec2}, which imply its vanishing:
\begin{itemize}
    \item The dynamical covariant derivative of $w_a$:
    \begin{align}\label{eq:nablaw}
    \nabla w_a
    &= \dot x^b \delta_b w_a - N^b{}_a w_b\\
    &= \dot x^b \partial_b w_a - 2 G^b \dot\partial_b w_a - N^b{}_a w_b \\
    &= p(t,s) w_a \dot t\,,
    \end{align}
    with
    \begin{align}
        p(t,s) = \frac{1}{2 h h''^2}\bigl[ -h''^2\partial_t h + (h'h''+hh''')\partial_t h'-hh''\partial_t h''\big]\,.
    \end{align}
    \item The dynamical covariant derivative of $T_{abc}$:
    \begin{align}\label{eq:nablaT}
        \nabla T_{abc} 
        &= \dot x^d \partial_d  T_{abc}  - 2 G^d \dot\partial_d  T_{abc} 
        - N^d{}_a T_{dbc} - N^d{}_b T_{dbc} - N^d{}_c T_{abd} \\
        &= q(t,s) \dot t T_{abc}\,,
    \end{align}
    with
    \begin{align}
        q(t,s) =  \frac{1}{2 sh h''^2}\Bigl[ -sh''^2\partial_t h+\bigl( h''(2h+sh')+shh'''\bigr)\partial_t h'-shh''\partial_t h''\Bigr]\,.
    \end{align}
\end{itemize}
The derivation of these results can most easily be done by considering the $3+1$ decomposed components of the equations, namely first $\nabla w_0$ and then $\nabla w_\alpha$, and first $T_{000}$ and then $T_{00\alpha}$, $T_{0\alpha\beta}$ as well as $T_{\alpha\beta\gamma}$. The index sums in \eqref{eq:nablaw} can also most easily be evaluated when they are expanded in a $3+1$ decomposition, by for example writing $\dot x^b \partial_b = \dot x^0 \partial_0 + \dot x^\beta\partial_\beta$. After doing so, one needs to plug in the explicit expressions \eqref{eq:geod-hom1}, \eqref{eq:geod-hom2} and \eqref{eq:nonlin-hom} of the spray, respectively of the nonlinear connection coefficients, which then, using the identity \eqref{eq:homw}, nicely factor as displayed in \eqref{eq:nablaw} and \eqref{eq:nablaT}.

\bibliography{UniCosmo}

\begin{thebibliography}{57}%
\makeatletter
\providecommand \@ifxundefined [1]{%
 \@ifx{#1\undefined}
}%
\providecommand \@ifnum [1]{%
 \ifnum #1\expandafter \@firstoftwo
 \else \expandafter \@secondoftwo
 \fi
}%
\providecommand \@ifx [1]{%
 \ifx #1\expandafter \@firstoftwo
 \else \expandafter \@secondoftwo
 \fi
}%
\providecommand \natexlab [1]{#1}%
\providecommand \enquote  [1]{``#1''}%
\providecommand \bibnamefont  [1]{#1}%
\providecommand \bibfnamefont [1]{#1}%
\providecommand \citenamefont [1]{#1}%
\providecommand \href@noop [0]{\@secondoftwo}%
\providecommand \href [0]{\begingroup \@sanitize@url \@href}%
\providecommand \@href[1]{\@@startlink{#1}\@@href}%
\providecommand \@@href[1]{\endgroup#1\@@endlink}%
\providecommand \@sanitize@url [0]{\catcode `\\12\catcode `\$12\catcode
  `\&12\catcode `\#12\catcode `\^12\catcode `\_12\catcode `\%12\relax}%
\providecommand \@@startlink[1]{}%
\providecommand \@@endlink[0]{}%
\providecommand \url  [0]{\begingroup\@sanitize@url \@url }%
\providecommand \@url [1]{\endgroup\@href {#1}{\urlprefix }}%
\providecommand \urlprefix  [0]{URL }%
\providecommand \Eprint [0]{\href }%
\providecommand \doibase [0]{https://doi.org/}%
\providecommand \selectlanguage [0]{\@gobble}%
\providecommand \bibinfo  [0]{\@secondoftwo}%
\providecommand \bibfield  [0]{\@secondoftwo}%
\providecommand \translation [1]{[#1]}%
\providecommand \BibitemOpen [0]{}%
\providecommand \bibitemStop [0]{}%
\providecommand \bibitemNoStop [0]{.\EOS\space}%
\providecommand \EOS [0]{\spacefactor3000\relax}%
\providecommand \BibitemShut  [1]{\csname bibitem#1\endcsname}%
\let\auto@bib@innerbib\@empty
\bibitem [{\citenamefont {Hohmann}\ \emph
  {et~al.}(2020{\natexlab{a}})\citenamefont {Hohmann}, \citenamefont
  {Pfeifer},\ and\ \citenamefont {Voicu}}]{Hohmann:2020yia}%
  \BibitemOpen
  \bibfield  {author} {\bibinfo {author} {\bibfnamefont {M.}~\bibnamefont
  {Hohmann}}, \bibinfo {author} {\bibfnamefont {C.}~\bibnamefont {Pfeifer}},\
  and\ \bibinfo {author} {\bibfnamefont {N.}~\bibnamefont {Voicu}},\ }\bibfield
   {title} {\bibinfo {title} {{The kinetic gas universe}},\ }\href
  {https://doi.org/10.1140/epjc/s10052-020-8391-y} {\bibfield  {journal}
  {\bibinfo  {journal} {Eur. Phys. J. C}\ }\textbf {\bibinfo {volume} {80}},\
  \bibinfo {pages} {809} (\bibinfo {year} {2020}{\natexlab{a}})},\ \Eprint
  {https://arxiv.org/abs/2005.13561} {arXiv:2005.13561 [gr-qc]} \BibitemShut
  {NoStop}%
\bibitem [{\citenamefont {Hohmann}\ \emph
  {et~al.}(2020{\natexlab{b}})\citenamefont {Hohmann}, \citenamefont
  {Pfeifer},\ and\ \citenamefont {Voicu}}]{Hohmann:2019sni}%
  \BibitemOpen
  \bibfield  {author} {\bibinfo {author} {\bibfnamefont {M.}~\bibnamefont
  {Hohmann}}, \bibinfo {author} {\bibfnamefont {C.}~\bibnamefont {Pfeifer}},\
  and\ \bibinfo {author} {\bibfnamefont {N.}~\bibnamefont {Voicu}},\ }\bibfield
   {title} {\bibinfo {title} {{Relativistic kinetic gases as direct sources of
  gravity}},\ }\href {https://doi.org/10.1103/PhysRevD.101.024062} {\bibfield
  {journal} {\bibinfo  {journal} {Phys. Rev. D}\ }\textbf {\bibinfo {volume}
  {101}},\ \bibinfo {pages} {024062} (\bibinfo {year} {2020}{\natexlab{b}})},\
  \Eprint {https://arxiv.org/abs/1910.14044} {arXiv:1910.14044 [gr-qc]}
  \BibitemShut {NoStop}%
\bibitem [{\citenamefont {Andreasson}(2011)}]{Andreasson:2011ng}%
  \BibitemOpen
  \bibfield  {author} {\bibinfo {author} {\bibfnamefont {H.}~\bibnamefont
  {Andreasson}},\ }\bibfield  {title} {\bibinfo {title} {{The Einstein-Vlasov
  System/Kinetic Theory}},\ }\href {https://doi.org/10.12942/lrr-2011-4}
  {\bibfield  {journal} {\bibinfo  {journal} {Living Rev. Rel.}\ }\textbf
  {\bibinfo {volume} {14}},\ \bibinfo {pages} {4} (\bibinfo {year} {2011})},\
  \Eprint {https://arxiv.org/abs/1106.1367} {arXiv:1106.1367 [gr-qc]}
  \BibitemShut {NoStop}%
\bibitem [{\citenamefont {Sarbach}\ and\ \citenamefont
  {Zannias}(2013)}]{Sarbach:2013fya}%
  \BibitemOpen
  \bibfield  {author} {\bibinfo {author} {\bibfnamefont {O.}~\bibnamefont
  {Sarbach}}\ and\ \bibinfo {author} {\bibfnamefont {T.}~\bibnamefont
  {Zannias}},\ }\bibfield  {title} {\bibinfo {title} {{Relativistic Kinetic
  Theory: An Introduction}},\ }\bibfield  {booktitle} {\emph {\bibinfo
  {booktitle} {{Proceedings, 9th Mexican School on Gravitation and Mathematical
  Physics: Cosmology for the XXI Century: Inflation, Dark Matter and Dark
  Energy (DGFM-SMF): Puerto Vallarta, Jalisco, Mexico, December 3-7, 2012}}},\
  }\href {https://doi.org/10.1063/1.4817035} {\bibfield  {journal} {\bibinfo
  {journal} {AIP Conf. Proc.}\ }\textbf {\bibinfo {volume} {1548}},\ \bibinfo
  {pages} {134} (\bibinfo {year} {2013})},\ \Eprint
  {https://arxiv.org/abs/1303.2899} {arXiv:1303.2899 [gr-qc]} \BibitemShut
  {NoStop}%
\bibitem [{\citenamefont {Ehlers}(2011)}]{Ehlers2011}%
  \BibitemOpen
  \bibfield  {author} {\bibinfo {author} {\bibfnamefont {J.}~\bibnamefont
  {Ehlers}},\ }\bibinfo {title} {General-relativistc kinetic theory of gases},\
  in\ \href {https://doi.org/10.1007/978-3-642-11099-3_4} {\emph {\bibinfo
  {booktitle} {Relativistic Fluid Dynamics}}},\ \bibinfo {editor} {edited by\
  \bibinfo {editor} {\bibfnamefont {C.}~\bibnamefont {Cattaneo}}}\ (\bibinfo
  {publisher} {Springer Berlin Heidelberg},\ \bibinfo {address} {Berlin,
  Heidelberg},\ \bibinfo {year} {2011})\ pp.\ \bibinfo {pages}
  {301--388}\BibitemShut {NoStop}%
\bibitem [{\citenamefont {Sarbach}\ and\ \citenamefont
  {Zannias}(2014)}]{Sarbach:2013uba}%
  \BibitemOpen
  \bibfield  {author} {\bibinfo {author} {\bibfnamefont {O.}~\bibnamefont
  {Sarbach}}\ and\ \bibinfo {author} {\bibfnamefont {T.}~\bibnamefont
  {Zannias}},\ }\bibfield  {title} {\bibinfo {title} {{The geometry of the
  tangent bundle and the relativistic kinetic theory of gases}},\ }\href
  {https://doi.org/10.1088/0264-9381/31/8/085013} {\bibfield  {journal}
  {\bibinfo  {journal} {Class. Quant. Grav.}\ }\textbf {\bibinfo {volume}
  {31}},\ \bibinfo {pages} {085013} (\bibinfo {year} {2014})},\ \Eprint
  {https://arxiv.org/abs/1309.2036} {arXiv:1309.2036 [gr-qc]} \BibitemShut
  {NoStop}%
\bibitem [{\citenamefont {Finsler}(1918)}]{Finsler}%
  \BibitemOpen
  \bibfield  {author} {\bibinfo {author} {\bibfnamefont {P.}~\bibnamefont
  {Finsler}},\ }\emph {\bibinfo {title} {\"{U}ber Kurven und Fl\"{a}chen in
  allgemeinen R\"{a}umen}},\ \href@noop {} {Ph.D. thesis},\ \bibinfo  {school}
  {Georg-August Universit\"{a}t zu G\"{o}ttingen} (\bibinfo {year}
  {1918})\BibitemShut {NoStop}%
\bibitem [{\citenamefont {Miron}\ and\ \citenamefont
  {Bucataru}(2007)}]{Bucataru}%
  \BibitemOpen
  \bibfield  {author} {\bibinfo {author} {\bibfnamefont {R.}~\bibnamefont
  {Miron}}\ and\ \bibinfo {author} {\bibfnamefont {I.}~\bibnamefont
  {Bucataru}},\ }\href@noop {} {\emph {\bibinfo {title} {Finsler Lagrange
  geometry}}}\ (\bibinfo  {publisher} {Editura Academiei Romane},\ \bibinfo
  {year} {2007})\BibitemShut {NoStop}%
\bibitem [{\citenamefont {Bao}\ \emph {et~al.}(2000)\citenamefont {Bao},
  \citenamefont {Chern},\ and\ \citenamefont {Shen}}]{BCS}%
  \BibitemOpen
  \bibfield  {author} {\bibinfo {author} {\bibfnamefont {D.}~\bibnamefont
  {Bao}}, \bibinfo {author} {\bibfnamefont {S.-S.}\ \bibnamefont {Chern}},\
  and\ \bibinfo {author} {\bibfnamefont {Z.}~\bibnamefont {Shen}},\ }\href@noop
  {} {\emph {\bibinfo {title} {{An Introduction to Finsler-Riemann
  Geometry}}}}\ (\bibinfo  {publisher} {Springer, New York},\ \bibinfo {year}
  {2000})\BibitemShut {NoStop}%
\bibitem [{\citenamefont {Pfeifer}(2019)}]{Pfeifer:2019wus}%
  \BibitemOpen
  \bibfield  {author} {\bibinfo {author} {\bibfnamefont {C.}~\bibnamefont
  {Pfeifer}},\ }\bibfield  {title} {\bibinfo {title} {{Finsler spacetime
  geometry in Physics}},\ }\href {https://doi.org/10.1142/S0219887819410044}
  {\bibfield  {journal} {\bibinfo  {journal} {Int. J. Geom. Meth. Mod. Phys.}\
  }\textbf {\bibinfo {volume} {16}},\ \bibinfo {pages} {1941004} (\bibinfo
  {year} {2019})},\ \Eprint {https://arxiv.org/abs/1903.10185}
  {arXiv:1903.10185 [gr-qc]} \BibitemShut {NoStop}%
\bibitem [{\citenamefont {Rutz}(1993)}]{Rutz}%
  \BibitemOpen
  \bibfield  {author} {\bibinfo {author} {\bibfnamefont {S.~F.}\ \bibnamefont
  {Rutz}},\ }\bibfield  {title} {\bibinfo {title} {{A Finsler generalisation of
  Einstein's vacuum field equations}},\ }\href@noop {} {\bibfield  {journal}
  {\bibinfo  {journal} {General Relativity and Gravitation}\ }\textbf {\bibinfo
  {volume} {25}},\ \bibinfo {pages} {1139} (\bibinfo {year}
  {1993})}\BibitemShut {NoStop}%
\bibitem [{\citenamefont {Triantafyllopoulos}\ \emph
  {et~al.}(2020)\citenamefont {Triantafyllopoulos}, \citenamefont
  {Kapsabelis},\ and\ \citenamefont {Stavrinos}}]{Triantafyllopoulos:2020ogl}%
  \BibitemOpen
  \bibfield  {author} {\bibinfo {author} {\bibfnamefont {A.}~\bibnamefont
  {Triantafyllopoulos}}, \bibinfo {author} {\bibfnamefont {E.}~\bibnamefont
  {Kapsabelis}},\ and\ \bibinfo {author} {\bibfnamefont {P.}~\bibnamefont
  {Stavrinos}},\ }\bibfield  {title} {\bibinfo {title} {{Gravitational Field on
  the Lorentz Tangent Bundle: Generalized Paths and Field Equations}},\ }\href
  {https://doi.org/10.1140/epjp/s13360-020-00570-x} {\bibfield  {journal}
  {\bibinfo  {journal} {Eur. Phys. J. Plus}\ }\textbf {\bibinfo {volume}
  {135}},\ \bibinfo {pages} {557} (\bibinfo {year} {2020})},\ \Eprint
  {https://arxiv.org/abs/2004.00356} {arXiv:2004.00356 [gr-qc]} \BibitemShut
  {NoStop}%
\bibitem [{\citenamefont {Javaloyes}\ \emph {et~al.}(2022)\citenamefont
  {Javaloyes}, \citenamefont {Sanchez},\ and\ \citenamefont
  {Villasenor}}]{Villasenor:2022-1}%
  \BibitemOpen
  \bibfield  {author} {\bibinfo {author} {\bibfnamefont {M.~A.}\ \bibnamefont
  {Javaloyes}}, \bibinfo {author} {\bibfnamefont {M.}~\bibnamefont {Sanchez}},\
  and\ \bibinfo {author} {\bibfnamefont {F.~F.}\ \bibnamefont {Villasenor}},\
  }\bibfield  {title} {\bibinfo {title} {{The Einstein–Hilbert–Palatini
  formalism in pseudo-Finsler geometry}},\ }\href
  {https://doi.org/10.4310/ATMP.2022.v26.n10.a5} {\bibfield  {journal}
  {\bibinfo  {journal} {Advances in Theoretical and Mathematical Physics}\
  }\textbf {\bibinfo {volume} {26}},\ \bibinfo {pages} {3563 – 3631}
  (\bibinfo {year} {2022})},\ \Eprint {https://arxiv.org/abs/2108.03197}
  {arXiv:2108.03197 [math.DG]} \BibitemShut {NoStop}%
\bibitem [{\citenamefont {Garcia-Parrado}\ and\ \citenamefont
  {Minguzzi}(2022)}]{Garcia-Parrado:2022ith}%
  \BibitemOpen
  \bibfield  {author} {\bibinfo {author} {\bibfnamefont {A.}~\bibnamefont
  {Garcia-Parrado}}\ and\ \bibinfo {author} {\bibfnamefont {E.}~\bibnamefont
  {Minguzzi}},\ }\bibfield  {title} {\bibinfo {title} {{An anisotropic gravity
  theory}},\ }\href {https://doi.org/10.1007/s10714-022-03039-7} {\bibfield
  {journal} {\bibinfo  {journal} {Gen. Rel. Grav.}\ }\textbf {\bibinfo {volume}
  {54}},\ \bibinfo {pages} {150} (\bibinfo {year} {2022})},\ \Eprint
  {https://arxiv.org/abs/2206.09653} {arXiv:2206.09653 [gr-qc]} \BibitemShut
  {NoStop}%
\bibitem [{\citenamefont {Pfeifer}\ and\ \citenamefont
  {Wohlfarth}(2011)}]{Pfeifer:2011tk}%
  \BibitemOpen
  \bibfield  {author} {\bibinfo {author} {\bibfnamefont {C.}~\bibnamefont
  {Pfeifer}}\ and\ \bibinfo {author} {\bibfnamefont {M.~N.~R.}\ \bibnamefont
  {Wohlfarth}},\ }\bibfield  {title} {\bibinfo {title} {{Causal structure and
  electrodynamics on Finsler spacetimes}},\ }\href@noop {} {\bibfield
  {journal} {\bibinfo  {journal} {Phys.Rev.}\ }\textbf {\bibinfo {volume}
  {D84}},\ \bibinfo {pages} {044039} (\bibinfo {year} {2011})},\ \Eprint
  {https://arxiv.org/abs/1104.1079} {arXiv:1104.1079 [gr-qc]} \BibitemShut
  {NoStop}%
\bibitem [{\citenamefont {Hohmann}\ \emph {et~al.}(2019)\citenamefont
  {Hohmann}, \citenamefont {Pfeifer},\ and\ \citenamefont
  {Voicu}}]{Hohmann_2019}%
  \BibitemOpen
  \bibfield  {author} {\bibinfo {author} {\bibfnamefont {M.}~\bibnamefont
  {Hohmann}}, \bibinfo {author} {\bibfnamefont {C.}~\bibnamefont {Pfeifer}},\
  and\ \bibinfo {author} {\bibfnamefont {N.}~\bibnamefont {Voicu}},\ }\bibfield
   {title} {\bibinfo {title} {{Finsler gravity action from variational
  completion}},\ }\href {https://doi.org/10.1103/PhysRevD.100.064035}
  {\bibfield  {journal} {\bibinfo  {journal} {Phys. Rev. D}\ }\textbf {\bibinfo
  {volume} {100}},\ \bibinfo {pages} {064035} (\bibinfo {year} {2019})},\
  \Eprint {https://arxiv.org/abs/1812.11161} {arXiv:1812.11161 [gr-qc]}
  \BibitemShut {NoStop}%
\bibitem [{\citenamefont {Kouretsis}\ \emph {et~al.}(2009)\citenamefont
  {Kouretsis}, \citenamefont {Stathakopoulos},\ and\ \citenamefont
  {Stavrinos}}]{Kouretsis:2008ha}%
  \BibitemOpen
  \bibfield  {author} {\bibinfo {author} {\bibfnamefont {A.}~\bibnamefont
  {Kouretsis}}, \bibinfo {author} {\bibfnamefont {M.}~\bibnamefont
  {Stathakopoulos}},\ and\ \bibinfo {author} {\bibfnamefont {P.}~\bibnamefont
  {Stavrinos}},\ }\bibfield  {title} {\bibinfo {title} {{The General Very
  Special Relativity in Finsler Cosmology}},\ }\href@noop {} {\bibfield
  {journal} {\bibinfo  {journal} {Phys.Rev.}\ }\textbf {\bibinfo {volume}
  {D79}},\ \bibinfo {pages} {104011} (\bibinfo {year} {2009})},\ \Eprint
  {https://arxiv.org/abs/0810.3267} {arXiv:0810.3267 [gr-qc]} \BibitemShut
  {NoStop}%
\bibitem [{\citenamefont {Mavromatos}\ \emph {et~al.}(2012)\citenamefont
  {Mavromatos}, \citenamefont {Mitsou}, \citenamefont {Sarkar},\ and\
  \citenamefont {Vergou}}]{Mavromatos:2010nk}%
  \BibitemOpen
  \bibfield  {author} {\bibinfo {author} {\bibfnamefont {N.~E.}\ \bibnamefont
  {Mavromatos}}, \bibinfo {author} {\bibfnamefont {V.~A.}\ \bibnamefont
  {Mitsou}}, \bibinfo {author} {\bibfnamefont {S.}~\bibnamefont {Sarkar}},\
  and\ \bibinfo {author} {\bibfnamefont {A.}~\bibnamefont {Vergou}},\
  }\bibfield  {title} {\bibinfo {title} {{Implications of a Stochastic
  Microscopic Finsler Cosmology}},\ }\href
  {https://doi.org/10.1140/epjc/s10052-012-1956-7} {\bibfield  {journal}
  {\bibinfo  {journal} {Eur. Phys. J.}\ }\textbf {\bibinfo {volume} {C72}},\
  \bibinfo {pages} {1956} (\bibinfo {year} {2012})},\ \Eprint
  {https://arxiv.org/abs/1012.4094} {arXiv:1012.4094 [hep-ph]} \BibitemShut
  {NoStop}%
\bibitem [{\citenamefont {Papagiannopoulos}\ \emph {et~al.}(2017)\citenamefont
  {Papagiannopoulos}, \citenamefont {Basilakos}, \citenamefont {Paliathanasis},
  \citenamefont {Savvidou},\ and\ \citenamefont
  {Stavrinos}}]{Papagiannopoulos:2017whb}%
  \BibitemOpen
  \bibfield  {author} {\bibinfo {author} {\bibfnamefont {G.}~\bibnamefont
  {Papagiannopoulos}}, \bibinfo {author} {\bibfnamefont {S.}~\bibnamefont
  {Basilakos}}, \bibinfo {author} {\bibfnamefont {A.}~\bibnamefont
  {Paliathanasis}}, \bibinfo {author} {\bibfnamefont {S.}~\bibnamefont
  {Savvidou}},\ and\ \bibinfo {author} {\bibfnamefont {P.~C.}\ \bibnamefont
  {Stavrinos}},\ }\bibfield  {title} {\bibinfo {title} {{Finsler–Randers
  cosmology: dynamical analysis and growth of matter perturbations}},\ }\href
  {https://doi.org/10.1088/1361-6382/aa8be1} {\bibfield  {journal} {\bibinfo
  {journal} {Class. Quant. Grav.}\ }\textbf {\bibinfo {volume} {34}},\ \bibinfo
  {pages} {225008} (\bibinfo {year} {2017})},\ \Eprint
  {https://arxiv.org/abs/1709.03748} {arXiv:1709.03748 [gr-qc]} \BibitemShut
  {NoStop}%
\bibitem [{\citenamefont {Li}\ \emph {et~al.}(2015)\citenamefont {Li},
  \citenamefont {Wang},\ and\ \citenamefont {Chang}}]{Li:2015sja}%
  \BibitemOpen
  \bibfield  {author} {\bibinfo {author} {\bibfnamefont {X.}~\bibnamefont
  {Li}}, \bibinfo {author} {\bibfnamefont {S.}~\bibnamefont {Wang}},\ and\
  \bibinfo {author} {\bibfnamefont {Z.}~\bibnamefont {Chang}},\ }\bibfield
  {title} {\bibinfo {title} {{Anisotropic inflation in the Finsler
  spacetime}},\ }\href {https://doi.org/10.1140/epjc/s10052-015-3468-8}
  {\bibfield  {journal} {\bibinfo  {journal} {Eur. Phys. J.}\ }\textbf
  {\bibinfo {volume} {C75}},\ \bibinfo {pages} {260} (\bibinfo {year}
  {2015})},\ \Eprint {https://arxiv.org/abs/1502.02256} {arXiv:1502.02256
  [gr-qc]} \BibitemShut {NoStop}%
\bibitem [{\citenamefont {Hohmann}\ and\ \citenamefont
  {Pfeifer}(2017)}]{Hohmann:2016pyt}%
  \BibitemOpen
  \bibfield  {author} {\bibinfo {author} {\bibfnamefont {M.}~\bibnamefont
  {Hohmann}}\ and\ \bibinfo {author} {\bibfnamefont {C.}~\bibnamefont
  {Pfeifer}},\ }\bibfield  {title} {\bibinfo {title} {{Geodesics and the
  magnitude-redshift relation on cosmologically symmetric Finsler
  spacetimes}},\ }\href {https://doi.org/10.1103/PhysRevD.95.104021} {\bibfield
   {journal} {\bibinfo  {journal} {Phys. Rev.}\ }\textbf {\bibinfo {volume}
  {D95}},\ \bibinfo {pages} {104021} (\bibinfo {year} {2017})},\ \Eprint
  {https://arxiv.org/abs/1612.08187} {arXiv:1612.08187 [gr-qc]} \BibitemShut
  {NoStop}%
\bibitem [{\citenamefont {Akrami}\ \emph {et~al.}(2021)\citenamefont {Akrami}
  \emph {et~al.}}]{CANTATA:2021asi}%
  \BibitemOpen
  \bibfield  {author} {\bibinfo {author} {\bibfnamefont {Y.}~\bibnamefont
  {Akrami}} \emph {et~al.} (\bibinfo {collaboration} {CANTATA}),\ }\href
  {https://doi.org/10.1007/978-3-030-83715-0} {\emph {\bibinfo {title}
  {{Modified Gravity and Cosmology. An Update by the CANTATA Network}}}},\
  edited by\ \bibinfo {editor} {\bibfnamefont {E.~N.}\ \bibnamefont
  {Saridakis}}, \bibinfo {editor} {\bibfnamefont {R.}~\bibnamefont {Lazkoz}},
  \bibinfo {editor} {\bibfnamefont {V.}~\bibnamefont {Salzano}}, \bibinfo
  {editor} {\bibfnamefont {P.}~\bibnamefont {Vargas~Moniz}}, \bibinfo {editor}
  {\bibfnamefont {S.}~\bibnamefont {Capozziello}}, \bibinfo {editor}
  {\bibfnamefont {J.}~\bibnamefont {Beltr\'an~Jim\'enez}}, \bibinfo {editor}
  {\bibfnamefont {M.}~\bibnamefont {De~Laurentis}},\ and\ \bibinfo {editor}
  {\bibfnamefont {G.~J.}\ \bibnamefont {Olmo}}\ (\bibinfo  {publisher}
  {Springer},\ \bibinfo {year} {2021})\ \Eprint
  {https://arxiv.org/abs/2105.12582} {arXiv:2105.12582 [gr-qc]} \BibitemShut
  {NoStop}%
\bibitem [{\citenamefont {Kapsabelis}\ \emph {et~al.}(2024)\citenamefont
  {Kapsabelis}, \citenamefont {Saridakis},\ and\ \citenamefont
  {Stavrinos}}]{Kapsabelis:2023khh}%
  \BibitemOpen
  \bibfield  {author} {\bibinfo {author} {\bibfnamefont {E.}~\bibnamefont
  {Kapsabelis}}, \bibinfo {author} {\bibfnamefont {E.~N.}\ \bibnamefont
  {Saridakis}},\ and\ \bibinfo {author} {\bibfnamefont {P.~C.}\ \bibnamefont
  {Stavrinos}},\ }\bibfield  {title} {\bibinfo {title}
  {{Finsler\textendash{}Randers\textendash{}Sasaki gravity and cosmology}},\
  }\href {https://doi.org/10.1140/epjc/s10052-024-12924-1} {\bibfield
  {journal} {\bibinfo  {journal} {Eur. Phys. J. C}\ }\textbf {\bibinfo {volume}
  {84}},\ \bibinfo {pages} {538} (\bibinfo {year} {2024})},\ \Eprint
  {https://arxiv.org/abs/2312.15552} {arXiv:2312.15552 [gr-qc]} \BibitemShut
  {NoStop}%
\bibitem [{\citenamefont {Stavrinos}\ and\ \citenamefont
  {Triantafyllopoulos}()}]{Stavrinos:2024noe}%
  \BibitemOpen
  \bibfield  {author} {\bibinfo {author} {\bibfnamefont {P.~C.}\ \bibnamefont
  {Stavrinos}}\ and\ \bibinfo {author} {\bibfnamefont {A.}~\bibnamefont
  {Triantafyllopoulos}},\ }\bibfield  {title} {\bibinfo {title} {{Cosmology
  Based on Finsler and Finsler-like Metric Structure of Gravitational Field}},\
  }\href@noop {} {\ }\Eprint {https://arxiv.org/abs/2405.19318}
  {arXiv:2405.19318 [gr-qc]} \BibitemShut {NoStop}%
\bibitem [{\citenamefont {Fuster}\ and\ \citenamefont
  {Pabst}(2016)}]{Fuster:2015tua}%
  \BibitemOpen
  \bibfield  {author} {\bibinfo {author} {\bibfnamefont {A.}~\bibnamefont
  {Fuster}}\ and\ \bibinfo {author} {\bibfnamefont {C.}~\bibnamefont {Pabst}},\
  }\bibfield  {title} {\bibinfo {title} {Finsler $pp$-waves},\ }\href
  {https://doi.org/10.1103/PhysRevD.94.104072} {\bibfield  {journal} {\bibinfo
  {journal} {Phys. Rev. D}\ }\textbf {\bibinfo {volume} {94}},\ \bibinfo
  {pages} {104072} (\bibinfo {year} {2016})},\ \Eprint
  {https://arxiv.org/abs/1510.03058} {arXiv:1510.03058} \BibitemShut {NoStop}%
\bibitem [{\citenamefont {Fuster}\ \emph {et~al.}(2018)\citenamefont {Fuster},
  \citenamefont {Pabst},\ and\ \citenamefont {Pfeifer}}]{Fuster:2018djw}%
  \BibitemOpen
  \bibfield  {author} {\bibinfo {author} {\bibfnamefont {A.}~\bibnamefont
  {Fuster}}, \bibinfo {author} {\bibfnamefont {C.}~\bibnamefont {Pabst}},\ and\
  \bibinfo {author} {\bibfnamefont {C.}~\bibnamefont {Pfeifer}},\ }\bibfield
  {title} {\bibinfo {title} {{Berwald spacetimes and very special
  relativity}},\ }\href {https://doi.org/10.1103/PhysRevD.98.084062} {\bibfield
   {journal} {\bibinfo  {journal} {Phys. Rev.}\ }\textbf {\bibinfo {volume}
  {D98}},\ \bibinfo {pages} {084062} (\bibinfo {year} {2018})},\ \Eprint
  {https://arxiv.org/abs/1804.09727} {arXiv:1804.09727 [gr-qc]} \BibitemShut
  {NoStop}%
\bibitem [{\citenamefont {Heefer}\ \emph {et~al.}(2021)\citenamefont {Heefer},
  \citenamefont {Pfeifer},\ and\ \citenamefont {Fuster}}]{Heefer_2021}%
  \BibitemOpen
  \bibfield  {author} {\bibinfo {author} {\bibfnamefont {S.}~\bibnamefont
  {Heefer}}, \bibinfo {author} {\bibfnamefont {C.}~\bibnamefont {Pfeifer}},\
  and\ \bibinfo {author} {\bibfnamefont {A.}~\bibnamefont {Fuster}},\
  }\bibfield  {title} {\bibinfo {title} {Randers
  $\mathrm{p}\mathrm{p}$-waves},\ }\href
  {https://doi.org/10.1103/PhysRevD.104.024007} {\bibfield  {journal} {\bibinfo
   {journal} {Phys. Rev. D}\ }\textbf {\bibinfo {volume} {104}},\ \bibinfo
  {pages} {024007} (\bibinfo {year} {2021})},\ \Eprint
  {https://arxiv.org/abs/2011.12969} {arXiv:2011.12969} \BibitemShut {NoStop}%
\bibitem [{\citenamefont {Heefer}\ and\ \citenamefont
  {Fuster}(2023)}]{Heefer_2023_Finsler_grav_waves}%
  \BibitemOpen
  \bibfield  {author} {\bibinfo {author} {\bibfnamefont {S.}~\bibnamefont
  {Heefer}}\ and\ \bibinfo {author} {\bibfnamefont {A.}~\bibnamefont
  {Fuster}},\ }\bibfield  {title} {\bibinfo {title} {Finsler gravitational
  waves of $(\alpha,\beta)$-type and their observational signature},\ }\href
  {https://doi.org/10.1088/1361-6382/acecce} {\bibfield  {journal} {\bibinfo
  {journal} {Class. Quantum Gravity}\ }\textbf {\bibinfo {volume} {40}},\
  \bibinfo {pages} {184002} (\bibinfo {year} {2023})},\ \Eprint
  {https://arxiv.org/abs/2302.08334} {arXiv:2302.08334} \BibitemShut {NoStop}%
\bibitem [{\citenamefont {Heefer}\ \emph
  {et~al.}(2023{\natexlab{a}})\citenamefont {Heefer}, \citenamefont {Pfeifer},
  \citenamefont {Reggio},\ and\ \citenamefont {Fuster}}]{Heefer:2023a}%
  \BibitemOpen
  \bibfield  {author} {\bibinfo {author} {\bibfnamefont {S.}~\bibnamefont
  {Heefer}}, \bibinfo {author} {\bibfnamefont {C.}~\bibnamefont {Pfeifer}},
  \bibinfo {author} {\bibfnamefont {A.}~\bibnamefont {Reggio}},\ and\ \bibinfo
  {author} {\bibfnamefont {A.}~\bibnamefont {Fuster}},\ }\bibfield  {title}
  {\bibinfo {title} {A cosmological unicorn solution to finsler gravity},\
  }\href {https://doi.org/10.1103/PhysRevD.108.064051} {\bibfield  {journal}
  {\bibinfo  {journal} {Phys. Rev. D}\ }\textbf {\bibinfo {volume} {108}},\
  \bibinfo {pages} {064051} (\bibinfo {year} {2023}{\natexlab{a}})}\BibitemShut
  {NoStop}%
\bibitem [{\citenamefont {Heefer}(2024{\natexlab{a}})}]{Heefer:2024kfi}%
  \BibitemOpen
  \bibfield  {author} {\bibinfo {author} {\bibfnamefont {S.}~\bibnamefont
  {Heefer}},\ }\emph {\bibinfo {title} {{Finsler Geometry, Spacetime \& Gravity
  -- From Metrizability of Berwald Spaces to Exact Vacuum Solutions in Finsler
  Gravity}}},\ \href@noop {} {Ph.D. thesis} (\bibinfo {year}
  {2024}{\natexlab{a}}),\ \Eprint {https://arxiv.org/abs/2404.09858}
  {arXiv:2404.09858 [gr-qc]} \BibitemShut {NoStop}%
\bibitem [{\citenamefont {Berwald}(1926)}]{Berwald1926}%
  \BibitemOpen
  \bibfield  {author} {\bibinfo {author} {\bibfnamefont {L.}~\bibnamefont
  {Berwald}},\ }\bibfield  {title} {\bibinfo {title} {Untersuchung der
  {K}r{\"u}mmung allgemeiner metrischer {R}{\"a}ume auf {G}rund des in ihnen
  herrschenden {P}arallelismus},\ }\href {https://doi.org/10.1007/BF01283825}
  {\bibfield  {journal} {\bibinfo  {journal} {Mathematische Zeitschrift}\
  }\textbf {\bibinfo {volume} {25}},\ \bibinfo {pages} {40} (\bibinfo {year}
  {1926})}\BibitemShut {NoStop}%
\bibitem [{\citenamefont {Szab\'o}(1981)}]{Szabo}%
  \BibitemOpen
  \bibfield  {author} {\bibinfo {author} {\bibfnamefont {Z.}~\bibnamefont
  {Szab\'o}},\ }\bibfield  {title} {\bibinfo {title} {{Positive definite
  Berwald spaces}},\ }\href@noop {} {\bibfield  {journal} {\bibinfo  {journal}
  {Tensor, New Series}\ }\textbf {\bibinfo {volume} {35}},\ \bibinfo {pages}
  {25} (\bibinfo {year} {1981})}\BibitemShut {NoStop}%
\bibitem [{\citenamefont {Crampin}(2014)}]{Crampin}%
  \BibitemOpen
  \bibfield  {author} {\bibinfo {author} {\bibfnamefont {M.}~\bibnamefont
  {Crampin}},\ }\bibfield  {title} {\bibinfo {title} {On the construction of
  riemannian metrics for berwald spaces by averaging},\ }\href@noop {}
  {\bibfield  {journal} {\bibinfo  {journal} {Houston Jour. Math.}\ }\textbf
  {\bibinfo {volume} {40}},\ \bibinfo {pages} {737} (\bibinfo {year}
  {2014})}\BibitemShut {NoStop}%
\bibitem [{\citenamefont {Hohmann}\ \emph
  {et~al.}(2020{\natexlab{c}})\citenamefont {Hohmann}, \citenamefont
  {Pfeifer},\ and\ \citenamefont {Voicu}}]{Hohmann:2020mgs}%
  \BibitemOpen
  \bibfield  {author} {\bibinfo {author} {\bibfnamefont {M.}~\bibnamefont
  {Hohmann}}, \bibinfo {author} {\bibfnamefont {C.}~\bibnamefont {Pfeifer}},\
  and\ \bibinfo {author} {\bibfnamefont {N.}~\bibnamefont {Voicu}},\ }\bibfield
   {title} {\bibinfo {title} {{Cosmological Finsler Spacetimes}},\ }\href
  {https://doi.org/10.3390/universe6050065} {\bibfield  {journal} {\bibinfo
  {journal} {Universe}\ }\textbf {\bibinfo {volume} {6}},\ \bibinfo {pages}
  {65} (\bibinfo {year} {2020}{\natexlab{c}})},\ \Eprint
  {https://arxiv.org/abs/2003.02299} {arXiv:2003.02299 [gr-qc]} \BibitemShut
  {NoStop}%
\bibitem [{\citenamefont {Fuster}\ \emph {et~al.}(2020)\citenamefont {Fuster},
  \citenamefont {Heefer}, \citenamefont {Pfeifer},\ and\ \citenamefont
  {Voicu}}]{Fuster:2020upk}%
  \BibitemOpen
  \bibfield  {author} {\bibinfo {author} {\bibfnamefont {A.}~\bibnamefont
  {Fuster}}, \bibinfo {author} {\bibfnamefont {S.}~\bibnamefont {Heefer}},
  \bibinfo {author} {\bibfnamefont {C.}~\bibnamefont {Pfeifer}},\ and\ \bibinfo
  {author} {\bibfnamefont {N.}~\bibnamefont {Voicu}},\ }\bibfield  {title}
  {\bibinfo {title} {{On the non metrizability of Berwald Finsler
  spacetimes}},\ }\href {https://doi.org/10.3390/universe6050064} {\bibfield
  {journal} {\bibinfo  {journal} {Universe}\ }\textbf {\bibinfo {volume} {6}},\
  \bibinfo {pages} {64} (\bibinfo {year} {2020})},\ \Eprint
  {https://arxiv.org/abs/2003.02300} {arXiv:2003.02300 [math.DG]} \BibitemShut
  {NoStop}%
\bibitem [{\citenamefont {Pfeifer}\ \emph {et~al.}(2021)\citenamefont
  {Pfeifer}, \citenamefont {Heefer},\ and\ \citenamefont
  {Fuster}}]{Pfeifer:2019tyy}%
  \BibitemOpen
  \bibfield  {author} {\bibinfo {author} {\bibfnamefont {C.}~\bibnamefont
  {Pfeifer}}, \bibinfo {author} {\bibfnamefont {S.}~\bibnamefont {Heefer}},\
  and\ \bibinfo {author} {\bibfnamefont {A.}~\bibnamefont {Fuster}},\
  }\bibfield  {title} {\bibinfo {title} {{Identifying Berwald Finsler
  geometries}},\ }\href {https://doi.org/10.1016/j.difgeo.2021.101817}
  {\bibfield  {journal} {\bibinfo  {journal} {Differ. Geom. Appl.}\ }\textbf
  {\bibinfo {volume} {79}},\ \bibinfo {pages} {101817} (\bibinfo {year}
  {2021})},\ \Eprint {https://arxiv.org/abs/1909.05284} {arXiv:1909.05284
  [math.DG]} \BibitemShut {NoStop}%
\bibitem [{\citenamefont {Cheraghchi}\ \emph {et~al.}(2023)\citenamefont
  {Cheraghchi}, \citenamefont {Pfeifer},\ and\ \citenamefont
  {Voicu}}]{Cheraghchi:2022zgv}%
  \BibitemOpen
  \bibfield  {author} {\bibinfo {author} {\bibfnamefont {S.}~\bibnamefont
  {Cheraghchi}}, \bibinfo {author} {\bibfnamefont {C.}~\bibnamefont
  {Pfeifer}},\ and\ \bibinfo {author} {\bibfnamefont {N.}~\bibnamefont
  {Voicu}},\ }\bibfield  {title} {\bibinfo {title} {{Four-dimensional
  SO(3)-spherically symmetric Berwald Finsler spaces}},\ }\href
  {https://doi.org/10.1142/S0219887823501906} {\bibfield  {journal} {\bibinfo
  {journal} {Int. J. Geom. Meth. Mod. Phys.}\ }\textbf {\bibinfo {volume}
  {20}},\ \bibinfo {pages} {2350190} (\bibinfo {year} {2023})},\ \Eprint
  {https://arxiv.org/abs/2212.08603} {arXiv:2212.08603 [math.DG]} \BibitemShut
  {NoStop}%
\bibitem [{\citenamefont {Voicu}\ \emph
  {et~al.}(2023{\natexlab{a}})\citenamefont {Voicu}, \citenamefont {Pfeifer},\
  and\ \citenamefont {Cheraghchi}}]{Voicu:2023fey}%
  \BibitemOpen
  \bibfield  {author} {\bibinfo {author} {\bibfnamefont {N.}~\bibnamefont
  {Voicu}}, \bibinfo {author} {\bibfnamefont {C.}~\bibnamefont {Pfeifer}},\
  and\ \bibinfo {author} {\bibfnamefont {S.}~\bibnamefont {Cheraghchi}},\
  }\bibfield  {title} {\bibinfo {title} {{Birkhoff theorem for Berwald-Finsler
  spacetimes}},\ }\href {https://doi.org/10.1103/PhysRevD.108.104060}
  {\bibfield  {journal} {\bibinfo  {journal} {Phys. Rev. D}\ }\textbf {\bibinfo
  {volume} {108}},\ \bibinfo {pages} {104060} (\bibinfo {year}
  {2023}{\natexlab{a}})},\ \Eprint {https://arxiv.org/abs/2306.07866}
  {arXiv:2306.07866 [math.DG]} \BibitemShut {NoStop}%
\bibitem [{\citenamefont {Heefer}\ \emph
  {et~al.}(2023{\natexlab{b}})\citenamefont {Heefer}, \citenamefont {Pfeifer},
  \citenamefont {van Voorthuizen},\ and\ \citenamefont
  {Fuster}}]{Heefer2023mKropNull}%
  \BibitemOpen
  \bibfield  {author} {\bibinfo {author} {\bibfnamefont {S.}~\bibnamefont
  {Heefer}}, \bibinfo {author} {\bibfnamefont {C.}~\bibnamefont {Pfeifer}},
  \bibinfo {author} {\bibfnamefont {J.}~\bibnamefont {van Voorthuizen}},\ and\
  \bibinfo {author} {\bibfnamefont {A.}~\bibnamefont {Fuster}},\ }\bibfield
  {title} {\bibinfo {title} {{On the metrizability of m-Kropina spaces with
  closed null one-form}},\ }\href
  {https://doi.org/https://doi.org/10.1063/5.0130523} {\bibfield  {journal}
  {\bibinfo  {journal} {J. Math. Phys.}\ }\textbf {\bibinfo {volume} {64}},\
  \bibinfo {pages} {022502} (\bibinfo {year} {2023}{\natexlab{b}})},\ \Eprint
  {https://arxiv.org/abs/2210.02718} {arXiv:2210.02718} \BibitemShut {NoStop}%
\bibitem [{\citenamefont
  {Heefer}(2024{\natexlab{b}})}]{heefer2024berwaldmkropinaspacesarbitrary}%
  \BibitemOpen
  \bibfield  {author} {\bibinfo {author} {\bibfnamefont {S.}~\bibnamefont
  {Heefer}},\ }\href {https://arxiv.org/abs/2407.00094} {\bibinfo {title}
  {Berwald $m$-{K}ropina spaces of arbitrary signature: Metrizability and
  {R}icci-flatness}} (\bibinfo {year} {2024}{\natexlab{b}}),\ \Eprint
  {https://arxiv.org/abs/2407.00094} {arXiv:2407.00094 [math.DG]} \BibitemShut
  {NoStop}%
\bibitem [{\citenamefont {Elgendi}(2021)}]{Elgendi2021a}%
  \BibitemOpen
  \bibfield  {author} {\bibinfo {author} {\bibfnamefont {S.}~\bibnamefont
  {Elgendi}},\ }\bibfield  {title} {\bibinfo {title} {Solutions for the
  {L}andsberg unicorn problem in {F}insler geometry},\ }\href
  {https://doi.org/https://doi.org/10.1016/j.geomphys.2020.103918} {\bibfield
  {journal} {\bibinfo  {journal} {Journal of Geometry and Physics}\ }\textbf
  {\bibinfo {volume} {159}},\ \bibinfo {pages} {103918} (\bibinfo {year}
  {2021})}\BibitemShut {NoStop}%
\bibitem [{\citenamefont {Elgendi}\ and\ \citenamefont
  {Kozma}(2021)}]{Elgendi2021b}%
  \BibitemOpen
  \bibfield  {author} {\bibinfo {author} {\bibfnamefont {S.~G.}\ \bibnamefont
  {Elgendi}}\ and\ \bibinfo {author} {\bibfnamefont {L.}~\bibnamefont
  {Kozma}},\ }\bibfield  {title} {\bibinfo {title} {$(\alpha,\beta)$-metrics
  satisfying the ${T}$-condition or the $\sigma {T}$-condition},\ }\href
  {https://doi.org/10.1007/s12220-020-00555-3} {\bibfield  {journal} {\bibinfo
  {journal} {The Journal of Geometric Analysis}\ }\textbf {\bibinfo {volume}
  {31}},\ \bibinfo {pages} {7866} (\bibinfo {year} {2021})}\BibitemShut
  {NoStop}%
\bibitem [{\citenamefont {Elgendi}(2020)}]{elgendi_2020}%
  \BibitemOpen
  \bibfield  {author} {\bibinfo {author} {\bibfnamefont {S.}~\bibnamefont
  {Elgendi}},\ }\bibfield  {title} {\bibinfo {title} {On the problem of
  non-{B}erwaldian {L}andsberg spaces},\ }\href
  {https://doi.org/10.1017/S000497271900128X} {\bibfield  {journal} {\bibinfo
  {journal} {Bulletin of the Australian Mathematical Society}\ }\textbf
  {\bibinfo {volume} {102}},\ \bibinfo {pages} {331} (\bibinfo {year}
  {2020})}\BibitemShut {NoStop}%
\bibitem [{\citenamefont {Bao}(2007)}]{Bao_unicorns}%
  \BibitemOpen
  \bibfield  {author} {\bibinfo {author} {\bibfnamefont {D.}~\bibnamefont
  {Bao}},\ }\bibfield  {title} {\bibinfo {title} {On two curvature-driven
  problems in {R}iemann-{F}insler geometry},\ }in\ \href@noop {} {\emph
  {\bibinfo {booktitle} {Finsler Geometry, Sapporo 2005: In Memory of Makoto
  Matsumoto}}},\ \bibinfo {series} {Adv. Stud. Pure Math.}, Vol.~\bibinfo
  {volume} {48}\ (\bibinfo  {publisher} {Mathematical Society of Japan},\
  \bibinfo {year} {2007})\ pp.\ \bibinfo {pages} {19--71}\BibitemShut {NoStop}%
\bibitem [{\citenamefont {Hohmann}\ \emph {et~al.}(2022)\citenamefont
  {Hohmann}, \citenamefont {Pfeifer},\ and\ \citenamefont
  {Voicu}}]{math-foundations}%
  \BibitemOpen
  \bibfield  {author} {\bibinfo {author} {\bibfnamefont {M.}~\bibnamefont
  {Hohmann}}, \bibinfo {author} {\bibfnamefont {C.}~\bibnamefont {Pfeifer}},\
  and\ \bibinfo {author} {\bibfnamefont {N.}~\bibnamefont {Voicu}},\ }\bibfield
   {title} {\bibinfo {title} {{Mathematical foundations for field theories on
  Finsler spacetimes}},\ }\href {https://doi.org/10.1063/5.0065944} {\bibfield
  {journal} {\bibinfo  {journal} {J. Math. Phys.}\ }\textbf {\bibinfo {volume}
  {63}},\ \bibinfo {pages} {032503} (\bibinfo {year} {2022})},\ \Eprint
  {https://arxiv.org/abs/2106.14965} {arXiv:2106.14965 [math-ph]} \BibitemShut
  {NoStop}%
\bibitem [{\citenamefont {Beem}(1970)}]{Beem}%
  \BibitemOpen
  \bibfield  {author} {\bibinfo {author} {\bibfnamefont {J.~K.}\ \bibnamefont
  {Beem}},\ }\bibfield  {title} {\bibinfo {title} {Indefinite {F}insler spaces
  and timelike spaces},\ }\href@noop {} {\bibfield  {journal} {\bibinfo
  {journal} {Can. J. Math.}\ }\textbf {\bibinfo {volume} {22}},\ \bibinfo
  {pages} {1035} (\bibinfo {year} {1970})}\BibitemShut {NoStop}%
\bibitem [{\citenamefont {Bernal}\ \emph {et~al.}(2020)\citenamefont {Bernal},
  \citenamefont {Javaloyes},\ and\ \citenamefont {S\'anchez}}]{Bernal2020}%
  \BibitemOpen
  \bibfield  {author} {\bibinfo {author} {\bibfnamefont {A.}~\bibnamefont
  {Bernal}}, \bibinfo {author} {\bibfnamefont {M.~A.}\ \bibnamefont
  {Javaloyes}},\ and\ \bibinfo {author} {\bibfnamefont {M.}~\bibnamefont
  {S\'anchez}},\ }\bibfield  {title} {\bibinfo {title} {{Foundations of Finsler
  Spacetimes from the Observers\textquoteright{} Viewpoint}},\ }\href
  {https://doi.org/10.3390/universe6040055} {\bibfield  {journal} {\bibinfo
  {journal} {Universe}\ }\textbf {\bibinfo {volume} {6}},\ \bibinfo {pages}
  {55} (\bibinfo {year} {2020})},\ \Eprint {https://arxiv.org/abs/2003.00455}
  {arXiv:2003.00455 [gr-qc]} \BibitemShut {NoStop}%
\bibitem [{\citenamefont {Lammerzahl}\ \emph {et~al.}(2012)\citenamefont
  {Lammerzahl}, \citenamefont {Perlick},\ and\ \citenamefont
  {Hasse}}]{Lammerzahl:2012kw}%
  \BibitemOpen
  \bibfield  {author} {\bibinfo {author} {\bibfnamefont {C.}~\bibnamefont
  {Lammerzahl}}, \bibinfo {author} {\bibfnamefont {V.}~\bibnamefont
  {Perlick}},\ and\ \bibinfo {author} {\bibfnamefont {W.}~\bibnamefont
  {Hasse}},\ }\bibfield  {title} {\bibinfo {title} {{Observable effects in a
  class of spherically symmetric static Finsler spacetimes}},\ }\href
  {https://doi.org/10.1103/PhysRevD.86.104042} {\bibfield  {journal} {\bibinfo
  {journal} {Phys. Rev.}\ }\textbf {\bibinfo {volume} {D86}},\ \bibinfo {pages}
  {104042} (\bibinfo {year} {2012})},\ \Eprint
  {https://arxiv.org/abs/1208.0619} {arXiv:1208.0619 [gr-qc]} \BibitemShut
  {NoStop}%
\bibitem [{\citenamefont {Hasse}\ and\ \citenamefont
  {Perlick}(2019)}]{Hasse:2019zqi}%
  \BibitemOpen
  \bibfield  {author} {\bibinfo {author} {\bibfnamefont {W.}~\bibnamefont
  {Hasse}}\ and\ \bibinfo {author} {\bibfnamefont {V.}~\bibnamefont
  {Perlick}},\ }\bibfield  {title} {\bibinfo {title} {{Redshift in Finsler
  spacetimes}},\ }\href {https://doi.org/10.1103/PhysRevD.100.024033}
  {\bibfield  {journal} {\bibinfo  {journal} {Phys. Rev.}\ }\textbf {\bibinfo
  {volume} {D100}},\ \bibinfo {pages} {024033} (\bibinfo {year} {2019})},\
  \Eprint {https://arxiv.org/abs/1904.08521} {arXiv:1904.08521 [gr-qc]}
  \BibitemShut {NoStop}%
\bibitem [{\citenamefont {Caponio}\ and\ \citenamefont
  {Masiello}(2020)}]{Caponio-Masiello}%
  \BibitemOpen
  \bibfield  {author} {\bibinfo {author} {\bibfnamefont {E.}~\bibnamefont
  {Caponio}}\ and\ \bibinfo {author} {\bibfnamefont {A.}~\bibnamefont
  {Masiello}},\ }\bibfield  {title} {\bibinfo {title} {{On the analyticity of
  static solutions of a field equation in Finsler gravity}},\ }\href
  {https://doi.org/10.3390/universe6040059} {\bibfield  {journal} {\bibinfo
  {journal} {Universe}\ }\textbf {\bibinfo {volume} {6}},\ \bibinfo {pages}
  {59} (\bibinfo {year} {2020})},\ \Eprint {https://arxiv.org/abs/2004.10613}
  {arXiv:2004.10613 [math.DG]} \BibitemShut {NoStop}%
\bibitem [{\citenamefont {Caponio}\ and\ \citenamefont
  {Stancarone}(2016)}]{Caponio-Stancarone}%
  \BibitemOpen
  \bibfield  {author} {\bibinfo {author} {\bibfnamefont {E.}~\bibnamefont
  {Caponio}}\ and\ \bibinfo {author} {\bibfnamefont {G.}~\bibnamefont
  {Stancarone}},\ }\bibfield  {title} {\bibinfo {title} {{Standard static
  Finsler spacetimes}},\ }\href {https://doi.org/10.1142/S0219887816500407}
  {\bibfield  {journal} {\bibinfo  {journal} {Int. J. Geom. Meth. Mod. Phys.}\
  }\textbf {\bibinfo {volume} {13}},\ \bibinfo {pages} {1650040} (\bibinfo
  {year} {2016})},\ \Eprint {https://arxiv.org/abs/1506.07451}
  {arXiv:1506.07451 [math.DG]} \BibitemShut {NoStop}%
\bibitem [{\citenamefont {Asanov}(2006{\natexlab{a}})}]{asanov_unicorns}%
  \BibitemOpen
  \bibfield  {author} {\bibinfo {author} {\bibfnamefont {G.}~\bibnamefont
  {Asanov}},\ }\bibfield  {title} {\bibinfo {title} {Finsleroid-{F}insler
  spaces of positive-definite and relativistic types},\ }\href
  {https://doi.org/https://doi.org/10.1016/S0034-4877(06)80053-4} {\bibfield
  {journal} {\bibinfo  {journal} {Rep. Math. Phys.}\ }\textbf {\bibinfo
  {volume} {58}},\ \bibinfo {pages} {275} (\bibinfo {year}
  {2006}{\natexlab{a}})}\BibitemShut {NoStop}%
\bibitem [{\citenamefont {Shen}(2009)}]{shen_unicorns}%
  \BibitemOpen
  \bibfield  {author} {\bibinfo {author} {\bibfnamefont {Z.}~\bibnamefont
  {Shen}},\ }\bibfield  {title} {\bibinfo {title} {On a class of {L}andsberg
  metrics in {F}insler geometry},\ }\href
  {https://doi.org/10.4153/CJM-2009-064-9} {\bibfield  {journal} {\bibinfo
  {journal} {Canadian Journal of Mathematics}\ }\textbf {\bibinfo {volume}
  {61}},\ \bibinfo {pages} {1357–1374} (\bibinfo {year} {2009})}\BibitemShut
  {NoStop}%
\bibitem [{\citenamefont {Weinberg}(1972)}]{Weinberg}%
  \BibitemOpen
  \bibfield  {author} {\bibinfo {author} {\bibfnamefont {S.}~\bibnamefont
  {Weinberg}},\ }\href@noop {} {\emph {\bibinfo {title} {{Gravitation and
  Cosmology}}}}\ (\bibinfo  {publisher} {John Wiley and Sons, New York},\
  \bibinfo {year} {1972})\BibitemShut {NoStop}%
\bibitem [{\citenamefont {Pfeifer}\ and\ \citenamefont
  {Wohlfarth}(2012)}]{Pfeifer:2011xi}%
  \BibitemOpen
  \bibfield  {author} {\bibinfo {author} {\bibfnamefont {C.}~\bibnamefont
  {Pfeifer}}\ and\ \bibinfo {author} {\bibfnamefont {M.~N.~R.}\ \bibnamefont
  {Wohlfarth}},\ }\bibfield  {title} {\bibinfo {title} {{Finsler geometric
  extension of Einstein gravity}},\ }\href@noop {} {\bibfield  {journal}
  {\bibinfo  {journal} {Phys.Rev.}\ }\textbf {\bibinfo {volume} {D85}},\
  \bibinfo {pages} {064009} (\bibinfo {year} {2012})},\ \Eprint
  {https://arxiv.org/abs/1112.5641} {arXiv:1112.5641 [gr-qc]} \BibitemShut
  {NoStop}%
\bibitem [{\citenamefont {Asanov}(2006{\natexlab{b}})}]{ASANOV2006275}%
  \BibitemOpen
  \bibfield  {author} {\bibinfo {author} {\bibfnamefont {G.}~\bibnamefont
  {Asanov}},\ }\bibfield  {title} {\bibinfo {title} {Finsleroid-finsler spaces
  of positive-definite and relativistic types},\ }\href
  {https://doi.org/https://doi.org/10.1016/S0034-4877(06)80053-4} {\bibfield
  {journal} {\bibinfo  {journal} {Reports on Mathematical Physics}\ }\textbf
  {\bibinfo {volume} {58}},\ \bibinfo {pages} {275} (\bibinfo {year}
  {2006}{\natexlab{b}})}\BibitemShut {NoStop}%
\bibitem [{\citenamefont {Voicu}\ \emph
  {et~al.}(2023{\natexlab{b}})\citenamefont {Voicu}, \citenamefont
  {Friedl-Sz\'asz}, \citenamefont {Popovici-Popescu},\ and\ \citenamefont
  {Pfeifer}}]{Voicu:2023zem}%
  \BibitemOpen
  \bibfield  {author} {\bibinfo {author} {\bibfnamefont {N.}~\bibnamefont
  {Voicu}}, \bibinfo {author} {\bibfnamefont {A.}~\bibnamefont
  {Friedl-Sz\'asz}}, \bibinfo {author} {\bibfnamefont {E.}~\bibnamefont
  {Popovici-Popescu}},\ and\ \bibinfo {author} {\bibfnamefont {C.}~\bibnamefont
  {Pfeifer}},\ }\bibfield  {title} {\bibinfo {title} {{The Finsler Spacetime
  Condition for (\ensuremath{\alpha},\ensuremath{\beta})-Metrics and Their
  Isometries}},\ }\href {https://doi.org/10.3390/universe9040198} {\bibfield
  {journal} {\bibinfo  {journal} {Universe}\ }\textbf {\bibinfo {volume} {9}},\
  \bibinfo {pages} {198} (\bibinfo {year} {2023}{\natexlab{b}})},\ \Eprint
  {https://arxiv.org/abs/2302.09937} {arXiv:2302.09937 [math.DG]} \BibitemShut
  {NoStop}%
\end{thebibliography}%

\end{document}